\newif\ifShowKeys
\ShowKeystrue

\documentclass[11pt]{article}
	\pdfoutput=1
\usepackage{etoolbox}
\makeatletter
\patchcmd{\l@section}
  {\addvspace{1.0em \@plus\p@}}
  {}
  {}{}
\makeatother

\usepackage{setspace}

\usepackage[usenames,dvipsnames]{xcolor}
\usepackage[setpagesize=false,pagebackref=false, linktocpage, bookmarksopen=true, colorlinks=true, linkcolor=Maroon,citecolor=Maroon,urlcolor=Maroon]{hyperref}
\usepackage[parsep]{collref}

\usepackage{amsmath, amssymb,amsthm}
\numberwithin{equation}{section}

\usepackage{bm,environ,mathrsfs,array,arydshln}
\usepackage{booktabs,float,slashed,hyperref}
\usepackage[mathcal]{euscript}
\usepackage{tensor} 						
\usepackage{mathabx}
\usepackage[vcentermath]{youngtab}
\usepackage{verbatim}

\usepackage{listings}
\usepackage{aurical}
\usepackage[T1]{fontenc}
\usepackage[nodayofweek]{date time}
\usepackage{graphicx,epsfig,epic}
\usepackage{tikz} 
\usetikzlibrary{arrows,decorations.pathreplacing,decorations.markings,snakes}
\usetikzlibrary{cd}

\usepackage{framed}						
\definecolor{shadecolor}{rgb}{0.9996078, 0.984314, 0.960784}

\usepackage{comment}

\allowdisplaybreaks

\definecolor{myred}{RGB}{233, 33, 45}

\newcommand{\bs}{\begin{shaded}}
\newcommand{\es}{\end{shaded}\noindent}
\def\ba#1\ea{\begin{align}#1\end{align}}		
\newcommand{\be}{\begin{equation}}
\newcommand{\ee}{\end{equation}}
\newcommand{\mc}{\mathcal }

\newcommand{\la}{\label}
\newcommand{\eps}{\varepsilon}

\newcommand{\lp}{\notag \\ & }

\newcommand{\cf}{\textit{cf.} }
\newcommand{\ie}{\textit{i.e.} }

\newcommand{\N}{\mathcal N}

\newcommand{\rf}[1]{(\ref{#1})}
\def \ci {\cite}
\def\foot{\footnote}
\def \del{\partial}
\def\ov{\over}
\def\G{\Gamma}
\def \four  {\tfrac{1}{4}}
\def \b  {\beta}

\def \ha {{1\ov 2}}
 \def \nb {n_{_\b}}
 
\def \no {\nonumber}
\def \ed {\end{document}}
\def \iffa {\iffalse}

\renewcommand{\l}{\lambda}

\DeclareMathOperator{\Tr}{Tr}

\DeclareMathOperator{\PE}{PE}
\DeclareMathOperator{\vol}{vol}

\newcommand{\mk}{\mathfrak}

\newcommand{\I}{\mathrm{I}}
\newcommand{\Isp}{\widehat{\I}}

\newcommand{\Z}{\mathcal Z}
\newcommand{\Zsp}{\widehat{\Z}}

\begin{document}




\begin{titlepage}
\begin{tabbing}
\hspace*{11.5cm} \=  \kill 
\>  Imperial-TP-AT-2023-06 \\
\> 
\end{tabbing}

\vspace*{15mm}
\begin{center}
{\Large\bf     Large $N$ expansion of   superconformal index  of $k=1$ ABJM  theory}\vskip 6pt
{\Large\bf   and  semiclassical   M5 brane partition function  }

\vspace*{10mm}

{\large M. Beccaria${}^{\,a,}$\footnote{ E-mail: \texttt{matteo.beccaria@le.infn.it}}   and 
A.A. Tseytlin$^{\,b,}$\footnote{\ Also at the Institute for Theoretical and Mathematical Physics (ITMP) of Moscow University   and Lebedev Institute, Moscow.}} 

\vspace*{4mm}
	
${}^a$ Universit\`a del Salento, Dipartimento di Matematica e Fisica \textit{Ennio De Giorgi},\\ 
		and I.N.F.N. - sezione di Lecce, Via Arnesano, I-73100 Lecce, Italy
			\vskip 0.1cm
${}^b$ Blackett Laboratory, Imperial College London SW7 2AZ, U.K.
			\vskip 0.1cm
\vspace*{0.8cm}
\end{center}

\begin{abstract}  
\noindent
It was shown in arXiv:2309.10786   that the 
leading  non-perturbative contribution  to the large $N$ expansion  of 
the superconformal index of the  (2,0)   6d theory 
(which describes   low-energy dynamics of  $N$ coincident M5 branes) 
 is   reproduced by the semiclassical partition function 
 of  quantum  M2  brane wrapped on $S^{1}\times S^{2}$ 
  in a twisted version of AdS$_{7}\times S^{4}$  background. 
  Here we  demonstrate  an  analogous   relation  for   the 
  leading  non-perturbative contribution  to  the large $N$ expansion  
  of  the superconformal index of  the  ${\cal N}=8$ supersymmetric level-one     $U(N) \times  U(N)$   
ABJM theory    (which describes    low-energy dynamics of  $N$ coincident M2 branes).
 The roles of  M2 and M5 branes  get effectively  interchanged.
 Namely,  the   large $N$  correction  to the  ABJM index 
 is    found to be given by the  semiclassical partition function 
 of   quantum  M5  brane  wrapped on $S^1 \times S^5$  in a twisted version of AdS$_{4}\times S^{7}$ 
 background.  This effectively  confirms  the suggestion  for the "M5 brane index"  made 
   in   arXiv:2007.05213   on the basis of  indirect  superconformal algebra considerations.
\end{abstract}
\vskip 0.5cm
	{
	}
\end{titlepage}

{
 \begin{spacing}{1.18}
\tableofcontents
\end{spacing}
}


\newpage

\def \tAdS {$\widetilde{\rm AdS}$}
 \def \rz {{\rm z}}
\def \zZ  {Z}
\def \tAdS {$\widetilde{\rm AdS}$}
 \def \rz {{\rm z}}
\def \zZ  {Z}
\def \rX  {{\rm X}}   \def \rY  {{\rm Y}}
\def \Mp  {\Pi} 
 \def \te {\textstyle} 
\def \hnabla {\hat \nabla} 
\def \bb {\bar \beta} 
\def \dd {{\rm d}}  \def \z  {\zeta}  \def \zz {{\cal Z}}  \def \half {{1\ov 2}}  \def \td {\tilde}
\def \te {\textstyle}  \def \ep {\eps}  \def \om {\omega} 
\def \bb {{\bar \beta}} \def \vp {\varphi} \def \a {\alpha}
\def \nbb {n_{_{\bb}}} \def \by {{\bar y}}
\def \half {{\frac{1}{2}}}
\def \pps {\phi} \def \wtd {\widetilde} 
\def \ga {\alpha} \def \rz {{\rm z}}
\def \GG  {{\rm G}}
\def \OO {{\cal O}}

\setcounter{footnote}{0}
\setcounter{section}{0}
\section{Introduction}

The superconformal index is the refined  Witten index for radially quantized superconformal theories \cite{Romelsberger:2005eg,Kinney:2005ej,Bhattacharya:2008zy}. 
It   has   schematic form  $\I= \Tr_{\rm BPS}[(-1)^{\rm F} q^H\prod_{a}u_{a}^{\mc C_{a}}]$ where the trace is restricted to BPS states annihilated by some chosen supercharge $Q$. 
$H$ and $\mc C_{a}$ are the Cartan generators of the symmetry superalgebra commuting with $Q$,  and $q$ and $u_{a}$ are free  parameters
(fugacities). 
For a superconformal theory on $\mathbb R^{d}$,  the index  
may be computed as a supersymmetric partition function 
on $S^{1}\times S^{d-1}$ \cite{Festuccia:2011ws,Dumitrescu:2012ha,Closset:2013vra}. This partition function 
 is defined by path integral 
with  fermions being periodic  in $S^{1}$ and including extra twists (rotations) of coordinates and/or fields that encode the presence of particular fugacities.

For the 6d superconformal $(2,0)$ theory
describing  the low-energy dynamics  of a stack of $N$ M5 branes, the corresponding  index $\I_{N}^{(2,0)}(q)$
in the Schur   limit \cite{Beem:2014kka} depends on a single fugacity $q$.
In  a  recent paper \cite{Beccaria:2023sph} it was   shown  that the 
leading large $N$  non-perturbative $\OO(q^N)$   contribution   to this index,
suggested in   \cite{Arai:2020uwd,Imamura:2021ytr} 
to  be related to the contribution of M2 brane  $S^{1}\times S^{2}$  instanton 
 in  AdS$_7\times S^4$,
  is   indeed reproduced by a semiclassical partition function 
 of   quantum  M2  brane in a twisted version of AdS$_{7}\times S^{4}$ 
 background. 

 Here we will demonstrate  that  the analogous  fact   is true for the leading large $N$  non-perturbative 
contribution to the 
superconformal index of  the  $\N=8$ supersymmetric $k=1$\    $U_{k}(N) \times  U_{-k}(N)$   
ABJM theory   \cite{Aharony:2008ug}  describing 
  low-energy dynamics of  $N$ coincident M2 branes,  
 with the roles of M2 and M5 branes effectively  interchanged.
 This  is consistent with the  expectation 
 that   supersymmetric partition function   of  the ABJM theory on $S^1 \times S^2$      should be
described on the  dual M-theory side by  a  quantum M2  brane   path integral 
 that may receive  non-perturbative large $N$  contributions from M5 brane instantons. 
 Namely,  the  leading large $N$ correction to the  ABJM index 
 will  be   found to be given by the  semiclassical partition function 
 of a  quantum  M5  brane  wrapped on $S^1 \times S^5$  in a twisted version of AdS$_{4}\times S^{7}$ 
 background.  This  will  confirm  an earlier  suggestion  of  \cite{Arai:2020uwd} based on  
superconformal algebra considerations. 



The computation  described below   demonstrates   that  a semiclassical  expansion of   quantum 
 M5 brane  path integral (around a  configuration   with a non-degenerate induced metric) 
 is well defined   and leads to   consistent  results.   
 It   provides an M5 brane   analog of  similar  computations   of semiclassical M2  brane path integrals 
 in several  recent works  \ci{Drukker:2020swu,Giombi:2023vzu,Beccaria:2023ujc,Seibold:2023zkz,Beccaria:2023sph,Drukker:2023jxp}.


Let us  first   recall the   result  for the index of the (2,0) theory.
For  $N=1$
 the  $(2,0)$ theory  is that of  a free 6d  tensor multiplet and the index
may be computed  using  representation theory \cite{Bhattacharya:2008zy}. 
For $N>1$, the absence of an intrinsic definition of the 
$(2,0)$ theory  may be  bypassed by  identifying its
  index on $S^{1}_{\beta}\times S^{5}$ ($\beta$ is the  length of the  circle)
with the  non-perturbative  supersymmetric 
partition function of the  maximally  supersymmetric 5d 
SYM theory  on $S^{5}$  (with  the coupling  proportional to $\beta$). 
This requires specific $R$-symmetry twists on $S^{1}_{\beta}\times S^{5}$
(corresponding to a particular  choice of Cartan  generators $\mc C_{i}$ in the definition of the index)  in order to 
preserve 16 real supersymmetries.
Using  this strategy, the index $\I_{N}^{(2,0)}$ was computed  by   supersymmetric 
 localization   in \cite{Kim:2012ava,Kim:2012qf}.
 
At large $N$, the  Schur
 index $\I_{N}^{(2,0)}(q)$   reduces to 
the generating function $\I_{\rm KK}^{(2,0)}(q)$ of  Kaluza-Klein BPS states of supergravity on AdS$_{7}\times S^{4}$.
Finite $N$ corrections  
 can be  organized as  an expansion in 
 powers of $q^{N}$  (with $q$-dependent  coefficients)  
\be
\la{1.1}
\I_{N}^{(2,0)}(q) = \I_{\rm KK}^{(2,0)}(q)\, \Big[1-\frac{q}{(1-q)^{2}}\,q^{N}+\mc O(q^{2N})\Big],\qquad \qquad q=e^{-\beta}\ .
\ee
Motivated by similar results for the index of  $\mc N=4$ $U(N)$  4d SYM theory  \cite{Imamura:2021ytr}
an alternative derivation of the  large $N$  correction  in  (\ref{1.1})   attributing it  to a  contribution of an 
M2  brane wrapped  on $S^{1}\times S^{2}$ in  AdS$_{7}\times S^{4}$  was   suggested  in \cite{Arai:2020uwd}.
In this approach, the  $q^{N}=e^{-\beta N}$  factor  originates  from the classical  value of the  wrapped 
M2 brane action while the  prefactor   should be  a superconformal index  counting   M2 brane   BPS fluctuations. 

This wrapped brane index was found   in  \cite{Arai:2020uwd} using 
 symmetry considerations,   without analyzing in detail the  world-volume theory.\foot{\la{fff1}
 Explicitly, the single M2  brane superconformal algebra   was  mapped to that of  the 3d  $k=1$ ABJM theory 
or of  the free theory of  $\N=8$ 3d scalar multiplet  (see, e.g.,   \cite{Gang:2011xp}).
The superalgebra mapping  implies  an analytic continuation relation between the   indices of the 
two 3d theories --  the wrapped M2  brane and the free $\N=8$ scalar multiplet  in flat space.  
This led the authors of \cite{Arai:2020uwd} to conjecture a remarkable general formula for 
the finite $N$ corrections to the index in terms of a sum over multiple  M2  wrappings,  reducing to (\ref{1.1}) for 
a  single wrapping.}
Instead,  ref. \cite{Beccaria:2023sph}   directly computed the  semiclassical 
 partition function  of the  M2  brane $S^{1}\times S^{2}$
  instanton     in  AdS$_{7}\times S^{4}$   reproducing 
  the $-\frac{q}{(1-q)^{2}}$   factor in \rf{1.1}  as the one-loop   contribution
  to the M2 brane path  integral. 


Below  we will consider the analog of  the large $N$ expansion \rf{1.1}   for the superconformal index 
of the $\N = 8$ $k=1$ $U(N)_{k}\times U(N)_{-k}$ ABJM  theory \cite{Aharony:2008ug}
dual to 11d M-theory on the AdS$_{4}\times S^{7}$ background.
Here the direct  definition of the ABJM  theory on $S^{1}_{\beta}\times S^{2}$ does not present a problem
and    the expression   for the corresponding index
can be found  \cite{Kim:2009wb}  using supersymmetric  localization  at   finite $N$.

The $N\to\infty$ limit of the ABJM index again counts the supergravity BPS states
on  AdS$_{4}\times S^{7}$.   For  finite $N$, the integral representation \cite{Kim:2009wb}   for 
 the index   can be used to extract its small $q$ expansion
but it is not suitable to  find its  non-perturbative expansion analogous to (\ref{1.1}).
The latter was   suggested in  \cite{Arai:2020uwd}   to  be related to  the contributions 
of  M5 branes wrapped   on $S^{1}_{\beta}\times S^{5}$ inside AdS$_{4}\times S^{7}$.
To  determine the corresponding  M5 brane index  ref. \cite{Arai:2020uwd}  suggested to 
use  an  analytic  continuation relation   to the index of a  single (2,0) tensor multiplet in flat space
that  happens to have 
 superconformal subalgebra compatible with the  supercharge  defining the index  which is 
isomorphic to the unbroken supersymmetry algebra on the M5 brane.

Starting with  the   expression for the M5 brane  index  suggested in   \cite{Arai:2020uwd} and taking an   unrefined  limit 
(i.e.  setting  all fugacities  apart from  single $q$   to be trivial) 
we found the  following analog  of \rf{1.1}\foot{Note that the counterpart  of $G_0$ 
 in the   large $N$ expansion of  (2,0) theory   index  in \rf{1.1} had a   much simpler rational form. 
 In \rf{1.1}  this    was   the 
 analytic continuation of the 
index of a free  3d $\N=8$ multiplet describing a single  M2  brane in  flat space,  
see   a discussion  in Appendix \ref{aC} and in particular \rf{x412}.}
\ba
\la{1.2}
&\I_{N}^{\rm ABJM}(q) = \I_{\rm KK}^{\rm ABJM}(q) \, \Big[1+\tfrac{1}{6}N^3\,q^{N} \big( G_0(q)  +  \mc O(N^{-1})\big)
  + \mc O(q^{2N})\Big], \\
 &  \qquad G_{0}(q) = -q\,\prod_{n=1}^{\infty}(1-q^{n})^{-7}\ , \qquad \ \   \ \ \ q = e^{-\ha \beta} \ . \la{1.22}
\ea
Note   that  the leading correction in  (\ref{1.2})  contains a peculiar $N^{3}$  enhancement factor that  had no counterpart
 in the 
 $(2,0)$ case in \rf{1.1}.\footnote{Such $N$-dependent factors  are common in  unrefined index expressions
  and are related to  the  "wall-crossing" effect \cite{Gaiotto:2021xce,Lee:2022vig,Beccaria:2023zjw}.
In general,    superconformal indices are protected against continuous deformations, but the coefficients of their 
 small $q$  expansions  can   jump across special surfaces in the space of global-symmetry fugacities.}

The aim  of the present  paper   is to  show that  the prefactor of the   leading $q^N$ 
term   in  (\ref{1.2})   can be  found  also 
 by the direct   one-loop $S^{1}_\beta \times S^{5}$ M5 brane instanton computation, 
   in full analogy with the M2  brane instanton  computation 
 in  the case of the (2,0) index \rf{1.1} in  \cite{Beccaria:2023sph}.
 
The  unrefined ABJM index   defined  by a combination of the radial  Hamiltonian and a 
rotation  generator  in one  plane of $\mathbb R^3$ 
 should be given by the   supersymmetric partition function of the 
ABJM  theory on  a  twisted  product $S^{1}_\b  \times \widetilde S^{2}$
with  an  isometric angle of $S^2$  being "mixed" with the coordinate $y$ of $S^1_\b $. 
The dual M-theory description should    be in terms of the AdS$_{4}\times S^{7}$ background 
with the  $S^{1}_\b \times \widetilde S^{2}$  boundary (here $y\equiv y + \beta$)\foot{\la{f7}This  may 
 be compared to the   twisted  AdS$_7\times S^4$ background 
 in  the $(2,0)$  theory case  considered in   \cite{Beccaria:2023sph} (see 
 also \cite{Bobev:2018ugk,Mezei:2018url,Gautason:2021vfc}):  
$ds^{2}_{11} = L'^{2} (dx^{2}+\sinh^{2}x\,dS_{5}+\cosh^{2}x\, dy^{2})+\tfrac{1}{4}L'^2 \big[dv^{2}+\cos^{2}v\,dS_{2}+\sin^{2}v\,(dz+idy)^{2}\big]$   supported by  $C_{3} = -\frac{1}{8}L'^{3} \cos^{2}v\,\vol_{S^{2}}\wedge(dz+idy)$.
}
\ba\la{1.3}
ds^2 =&\tfrac{1}{4}L^{2}\, ( dx^2 +  \sinh^2 x\, d\widetilde{S}_{2} +   \cosh^2 x\,  dy^2)+L^{2}
( dv^2 + \cos^2 v   \, dS_{5} + \sin^2 v \, dz^2 )\ , \\
\la{1.4}
d\widetilde{S}_{2} =& d\vartheta^{2}+\sin^{2}\vartheta\,(d\varphi+i \, dy)^{2},\ \ \ \ \ \ \
 \  \qquad   C_{3} = -  \tfrac{i}{8}  L^3 \, \sinh^{3}x\,   dy \wedge \vol_{S^{2}} \ . 
\ea
 Below we  will   compute the  semiclassical path integral for  the M5 brane  wrapped 
 on $S^{1}_{\beta}\subset$ AdS$_4$    and $ S^{5}\subset S^7$.
{ As   the effective M5 brane tension is given by 
 ${\rm T_5} = L^6 T_5= {1\ov \pi^3} N$,   the  $q^N$ term  in \rf{1.2} 
 will   come from   the classical action of  the 
 wrapped  M5 brane. 
 Its prefactor   will   be given   by  the one-loop M5 brane partition function, with 
 the   $N^3\sim (\sqrt{\rm T_5})^6$ factor  being  due to the  zero-mode  contribution
 and  $G_0(q)$  originating  from  the  combination of  determinants of the 
   M5 brane  quadratic  fluctuation operators.  
 
 The set of  M5  brane fluctuations in the  static gauge   will  generalize  the  one of the 
 6d $(2,0)$ tensor multiplet  in flat target space, i.e. it will 
  contain 3  scalars of AdS$_4$, 2 scalars of $S^7$, 
  the rank 2 antisymmetric tensor with self-dual field  strength and 4  Majorana-Weyl   6d fermions
  propagating on $S^1_\b  \times S^5$. 
   We will separately compute the  contributions  of all of these fields 
   that  will  thus involve all  fluctuations, not just BPS one.     Combined together  
    the  resulting  supersymmetric partition function   will effectively reproduce the 
   expression for the  supersymmetric M5 brane index  conjectured in \cite{Arai:2020uwd}. 
     This  result  will  crucially  depend  on the special 
   structure of the  supersymmetric M5 brane action  in  the background \rf{1.3},\rf{1.4} 
    (e.g. the coupling to $C_3$   background   via  the "dual"  $C_6$  potential and 
   fermionic  covariant derivatives, etc.).

We shall start in section 2   with a review of the structure of the  large $N$ expansion  of the 
superconformal index   of  the $k=1$   ABJM theory. 
In section 3  we will consider  the supersymmetric M5 brane path integral in
semiclassical expansion near $S^1 \times S^5$ solution  in 
 the background \rf{1.3},\rf{1.4}. We will  find  the quadratic fluctuation actions for all M5 brane fluctuations
 and present the results for the corresponding one-loop  contributions to partition function. 
 
 The  combined  expression for the one-loop  M5 brane  supersymmetric partition function or free energy 
 will be given in section 4.  We will  demonstrate that  it matches the expression for the $N^3 G_0(q)$ prefactor of  $q^N$ in 
 the unrefined ABJM index \rf{1.2}.   In section 4.3
 we will 
  also  consider a generalization to the case of non-trivial  fugacities $u_a$  implying the presence of extra twist angles in $S^7$ 
 part of \rf{1.3}. We will  find again a precise
 matching  between the corresponding 
  M5 brane  partition  and  the expression for the  M5 brane index  suggested in \cite{Arai:2020uwd}.
 
 In section  5 we will show how the   general expressions for the  fluctuation determinants 
  of the  fields  of a  single   (2,0)   multiplet on $S^1 \times S^5$ 
 found in section  3   can be used  to  directly compute the  corresponding  supersymmetric partition function or  the 
 $N=1$ (2,0) 
 superconformal index.  

In Appendix A we shall  make some remarks  about brane instanton (or "giant graviton")  expansion of superconformal index on the example of   $\N=4$ SYM theory.   In Appendix B  we shall  review the  definition of the superconformal index of (2,0)  theory.
In Appendix C,   starting with the  general expression   for the M5 brane index suggested  in  \cite{Arai:2020uwd},
  we   will   show that taking  the  unrefined limit  of the   corresponding  large $N$ expansion 
   of the ABJM    index  leads to  the expression in \rf{1.2}. 
  In  Appendix D  we will derive  the  free energy  of a conformal  scalar field on $S^1 \times S^5$ with an  extra   shift 
  of  the  $S^1$ mode number  that was used in section 4. 
  Appendix E will contain  details  of the computation discussed in section 4.3.

\section{Superconformal index  of $k=1$   ABJM theory  at large $N$  }

The superconformal algebra of the 3d $\N = 8$  supersymmetric 
 $U_k (N)\times U_{-k} (N)$ ABJM theory with Chern-Simons level $k = 1$ 
   is $\mk{osp}(8|4)$. Its 
  bosonic subalgebra  $\mk{so}(2,3)\oplus \mk{so}(8)$   is the sum of  the 3d  conformal algebra and the 
   $R$-symmetry   algebra 
   with six Cartan generators\be
{\mc C}_I = (H, \ J_{12}, \  R_{12}, \ R_{34}, \ R_{56}, \ R_{78}).
\ee
The superconformal index is defined in terms of the  supercharge $Q$   satisfying 
$[\mc C_{I},Q]=\frac{1}{2}\sigma_{I}\,Q$ with ${\sigma}_I=(1,-1,1,1,1,1)$\   ($I=1, \dots, 6$).
The subalgebra commuting with $Q$ is $\mk{osp}(6|2)\oplus \mk{u}(1)_{\Delta}$, where $\mk{osp}(6|2)$ has bosonic
algebra $\mk{sl}(2,\mathbb R)\oplus \mk{so}(6)$ and the central factor $\mk{u}(1)_{\Delta}$ is generated by 
\be
\la{2.2}
\Delta = \{Q,Q^{\dagger}\} = H-J_{12}-\tfrac{1}{2}(R_{12}+R_{34}+R_{56}+R_{78}).
\ee
The  index is   given by 
\be
\la{2.3}
\I_{N}^{\rm ABJM}(q, \bm{u}) = \mathop{\Tr}_{\Delta=0}\Big[(-1)^{\rm F}\, q^{2(H+J_{12})}\, u_{1}^{R_{12}}\, u_{2}^{R_{34}}\, u_{3}^{R_{56}}\, u_{4}^{R_{78}}\Big],\qquad u_{1}u_{2}u_{3}u_{4}=1,
\ee
where the trace is restricted to gauge singlet states with $\Delta=0$
and $u_a$ are $R$-symmetry fugacities. 
 An explicit integral representation of (\ref{2.3})
 was  obtained in \cite{Kim:2009wb} by  using supersymmetric localization.
It is valid also for  the $\mc N=6$  supersymmetric  ABJM theory with generic   level $k$,
and  for $N>1$ includes the contribution of monopole operators.

We will denote by $\Isp$ the 
 single-particle index  corresponding  to  the 
index $\I$, i.e. related to it   by  the plethystic exponential (here $\bm{x}$   stands for all  fugacities  including $q$)
\be \la{2.4}
\I(\bm{x}) = \PE[\Isp(\bm{x})] \equiv \exp\sum_{m=1}^{\infty}\frac{1}{m}\, \Isp(\bm{x}^{m})\ .
\ee
At leading order in  large $N$,
 the superconformal index  coincides with the one that  counts  the  BPS  states of 
 supergravity  in AdS$_{4}\times S^{7}$  (i.e. KK  states  of 11d supergravity compactified on $S^7$) \cite{Kim:2009wb}.
  It  has the following explicit expression  \cite{Bhattacharya:2008zy} 
\ba
\I_{N}^{\rm ABJM}(q,\bm{u}) &  \ \stackrel{N\to\infty}{\to}\  \I_{\rm KK}(q,\bm{u})= \PE[ \Isp_{\rm KK}(q,\bm{u})], \notag \\
\la{2.5}
\Isp_{\rm KK}(q,\bm{u}) =& \frac{1}{(1-q^{4})^{2}}\,\prod_{a=1}^{4}\frac{1-q^{3}\,u_{a}^{-1}}{1-q\, u_{a}}-\frac{1-q^{4}+q^{8}}{(1-q^{4})^{4}}.
\ea
To understand the structure of finite $N$ corrections, it is useful to examine the  
small $q$ expansion of the index at low  values of $N$. From the integral representation in  \cite{Kim:2009wb}  one finds 
(here for simplicity we set  all  $u_a= 1$)\ba
\la{2.6}
\I^{\rm ABJM}_{N=1}(q) =& \bm{1}+\bm{4}\,q+10\,q^{2}+16\,q^{3}+19\,q^{4}+20\,q^{5}+\cdots, \notag \\ 
\I^{\rm ABJM}_{N=2}(q) =& \bm{1}+\bm{4}\,q+\bm{20}\,q^{2}+56\,q^{3}+139\,q^{4}+260\,q^{5}+\cdots, \notag \\ 
\I^{\rm ABJM}_{N=3}(q) =& \bm{1}+\bm{4}\,q+\bm{20}\,q^{2}+\bm{76}\,q^{3}+239\,q^{4}+644\,q^{5}+\cdots . 
\ea
Here  $\I^{\rm ABJM}_{N=1}(q)  = \PE[\frac{4q}{1+q^{2}}]$  (see, e.g.,  \cite{Kim:2009wb,Gang:2011xp}). In (\ref{2.6})
we  highlighted  the expansion coefficients that  remain  stable under further increase of $N$ and whose contribution 
reproduces the KK index  in (\ref{2.5})
\be
 \I_{\rm KK}(q, \bm{u}=1) = 1+4q+20q^{2}+76q^{3}+274 q^{4}+900q^{5}+2826q^{6}+8400 q^{8}+\cdots .
\ee
The  structure  of   large $N$ 
 expansion    implied by  (\ref{2.6})  and its generalization to non-trivial $\bm u$
 is the following 
\ba
\la{2.8}
\I_{N}^{\rm ABJM}(q, \bm{u}) =& \I_{\rm KK}(q, \bm{u})\, \Big[1+q^{N}\, \delta\I^{(1)}_{N}(q,\bm{u})+\mc O(q^{2N+6}) \Big]
\ . 
 \ea
Here 
$\delta\I^{(1)}_{N}(q,\bm{u})$ is the first term in an 
infinite series of non-perturbative corrections  $\sum_n q^{nN}\delta\I^{(n)}(q,\bm{u})$
(\cf Appendix \ref{app:brane-exp}).

It was suggested in  \cite{Arai:2020uwd} that  in  the dual M-theory  description of the ABJM  theory 
the large $N$ corrections to the index   may  be  interpreted as  originating from 
 contributions of   M5 brane instantons in AdS$_4 \times S^7$. 
In particular,  the $q^{nN}$  term  may come
from M5 brane wrapped    $n$ times on $S^{1}\times S^{5}$  part of AdS$_4 \times S^7$.\foot{The corresponding classical  action then  produces the factor of $q^{n N}\, u_{a}^{N}$    taking into account that  the effective 
M5 brane tension is proportional to $N$, see 
 below.} 
Ref.  \cite{Arai:2020uwd}   conjectured 
 the  following expression  for  $\delta\I^{(1)}_{N}(q, \bm{u})$
\ba
\la{2.9}
\delta\I^{(1)}_{N}(q, \bm{u}) =& \sum_{a=1}^{4}u_{a}^{N}\,  \I^{\rm M5}_{a}(q,\bm{u}) \ , \ \ \ \ \ \ \ \ \ 
  \I^{\rm M5}_{a}(q,\bm{u})] =  \PE[ \Isp^{\rm M5}_{a}(q,\bm{u})],
\ea
where 
the suitably defined
 "M5 brane index" $\I^{\rm M5}_{a}$   should be     counting   BPS   fluctuations  of  
an 
 M5 brane wrapped on $S^{1}\times S^{5}$ in AdS$_4 \times S^7$  with the 
embedding  of $S^{5}\subset S^{7}$   corresponding    to  the choice of  one plane 
  $\rz_{a}=0$ in  the  definition    $\sum_{c=1}^{4}|\rz_{c}|^{2}=1$  of $S^7 \subset \mathbb C^4$.\foot{For the ABJM theory at level $k\ge 2$, the dual geometry is the orbifold $AdS_{4}\times S^{7}/\mathbb Z_{k}$.
For each of the four wrapping configurations $\rz_{a}=0$  ($a=1, \dots, 4$) 
 the M5 brane world volume is $S^{1}\times (S^{5}/\mathbb Z_{k})$.
M5 branes have a topological wrapping number $b\in \mathbb Z_{k}$, corresponding to the non-trivial homology
$H_{5}(S^{7}/\mathbb Z_{k})=\mathbb Z_{k}$, 
and the index can be computed in each sector with given $b$.
The  $b=0$  case corresponds to  the $U(1)_{k}\times U(1)_{-k}$ ABJM  theory. 
Finite $N$ corrections to the index start  with multiple wound branes for $k=2,3$ and presumably for all $k\ge 2$ \cite{Arai:2020uwd}. 
Wrapped branes with $b\neq 0$ correspond  to baryonic operators in the gauge theory charged under a $\mathbb Z_{k}$ subgroup of the gauge symmetry \cite{Bergman:2020ifi,Tachikawa:2019dvq}.
}  

The explicit computation of  the index $\I^{\rm M5}_{a}$ in \rf{2.9} requires the analysis of 
the  set of  fluctuation modes on the wrapped M5 brane 
  which was  not   carried out in full  in \cite{Arai:2020uwd}. 
Instead,   ref.  \cite{Arai:2020uwd}  suggested to 
 exploit   the isomorphism\footnote{
 {Details of the construction are similar to the one in 
 the M2  brane case discussed in footnote \ref{fff1}.}} between the  superconformal  algebras of the two 6d models   -- 
   the  quadratic fluctuation  action  of  the wrapped M5 brane  in  AdS$_4 \times S^7$  space and   free 
   (2,0) tensor multiplet  defined on $S^{1}\times S^{5}$.  This allowed 
 to find $\I^{\rm M5}_{a}$  from the   known  expression  for  the superconformal index of the 
 $N=1$  (2,0)  multiplet  (see Appendix \ref{aC})
 \be \la{222}
 \I^{(2,0)}_{N=1}(q', \bm{y}, u') = \mathop{\Tr}_{\Delta=0}\Big[(-1)^{\rm F}\,  q'^{H+\frac{1}{3}(J_{12}+J_{34}+J_{56})}\,
 y_{1}^{J_{12}}\, y_{2}^{J_{34}}\, y_{3}^{J_{56}}\, u'^{R_{12}-R_{34}}\Big],\qquad y_{1}y_{2}y_{3}=1 \ .\ee
 Here $J_{ij}$ are  the  generators  of the $SO(6)$ rotation group  of  the 6d space  on which 
 the (2,0)  multiplet  is  defined 
  and $R_{ab}$ are  the Cartan generators of the $SO(5)$ R-symmetry group. 
 The  reason  why  the two functions   $\I^{\rm M5}_{a}$   and  $\I^{(2,0)}_{N=1}$    
 should be related 
should be supersymmetry  (as count of BPS states  should require only group theory   information).
This  argument   appears to be rather   heuristic, given that  details of $\I^{\rm M5}_{a}$ 
should   depend on a  particular  structure 
of the  supersymmetric  M5 brane action in curved  AdS$_4 \times S^7$ background 
   (with non-trivial  coupling to the  4-form flux, etc.).  

The isomorphism   between the  generators of the  two superconformal algebras 
 determines the identification of  the two sets of  fugacities 
   needed to get $\I^{\rm M5}_{a} (q, u_1, u_2, u_3,u_4)$ 
from the $N=1$ (2,0)  index  in \rf{222}. 
For the $\I^{\rm M5}_{1}$ term in \rf{2.9}   that  distinguishes  the $u_1$  fugacity    this relation   reads
\be
\la{2.10}
(q', \ y_{1},\ y_{2},\ y_{3},\ u') = (q^{\frac{3}{4}}u_{1}^{-\frac{1}{4}}, \ u_{2}u_{1}^{\frac{1}{3}}, \ 
u_{3}u_{1}^{\frac{1}{3}}, \ u_{4}u_{1}^{\frac{1}{3}}, \ q^{-\frac{5}{2}}u_{1}^{-\frac{1}{2}}) \ .
\ee
As a result,   one gets the following expression for the  corresponding single-particle M5 brane   index 
\cite{Arai:2020uwd} 
\be
\la{2.11}
\Isp^{\rm M5}_{1}(q,\bm{u}) = \frac{q^{-1}u_{1}^{-1}-q^{2}u_{1}^{-1}(u_{2}^{-1}+u_{3}^{-1}+u_{4}^{-1})+q^{3}u_{1}^{-1}+q^{4}}{(1-q u_{2})(1-q u_{3})(1-q u_{4})}.
\ee
From   (\ref{2.3})   we can identify $q $  here with $ e^{-{1\ov 2} \beta}$ where $\beta$ is the length of the 
 circle in  the 3d space $S^{1}_{\beta}\times S^{2}$ on which ABJM theory is defined. 
$\Isp^{\rm M5}_{a} $ terms with $a=2,3,4$  in the sum in \rf{2.9}  
 are then  obtained from (\ref{2.11}) by  the permutations of $u_1, u_2, u_3, u_4$.

As  we  show in Appendix \ref{app:finiteN},  in the  simplest "unrefined"  case    of  $\bm{u}\to  1$ 
the leading term  $\delta\I^{(1)}(q, \bm{u})$ in \rf{2.9}  has the  following explicit   form 
\be
\la{2.12}
\delta \I^{(1)}_{N}(q) \equiv \delta \I^{(1)}_{N}(q,\bm{u}= 1)= \frac{1}{6}N^{3}G_{0}(q) + N^{2}G_{1}(q)+NG_{2}(q)+G_{3}(q)\ ,
\ee
where  
\be
\la{2.13}
G_{0}(q) = -q\,	\prod_{n=1}^{\infty}\frac{1}{(1-q^{n})^{7}} = -q\,\PE\Big[\frac{7q}{1-q}\Big]\ .
\ee
Below,  in sections 3 and 4,   we will reproduce \rf{2.12},\rf{2.13}  by the
 direct computation of the  one-loop  supersymmetric   partition function of the 
M5 brane instanton  in the relevant  twisted version of the AdS$_4 \times S^7$   background, confirming the conjecture of   \cite{Arai:2020uwd}.
In section 4.3   we  will also discuss  a generalization to the  case  with non-trivial $u_a$.

\section{Semiclassical expansion of M5 brane path integral}
\la{sec:Mtheory}


Our   aim   will  be to show 
 that  the leading $q^{N}\, \delta\I^{(1)}_{N}(q,\bm{u})$ term in  the ABJM index  in \rf{2.8} is reproduced 
by the  path integral contribution of  an   M5 brane $ S^1_\beta\times S^5$    instanton in a particular  twisted version of the 
AdS$_4 \times S^7$    M-theory background   dual  to the   $k=1$ ABJM theory. 

Here  we  shall   focus on the simplest case  of  the index \rf{2.3} with  all $u_a=1$, i.e. 
$\Tr[(-1)^{\rm F}e^{-\beta(H+J_{12})}]$ that should be given  by  the    
supersymmetric partition function  of the ABJM theory on $S^{1}_{\beta}\times \widetilde S^{2}$
(with periodic fermions  and  an extra   shift  of the $\widetilde S^2$ angle 
 related to the  shift of $H$   by  the $J_{12}$ rotational generator, cf.  \cite{Aharony:2021zkr}).
The dual M-theory  background  \tAdS$_4\times S^7$  is  then the following  "twisted"   analog of
the Euclidean  AdS$_4 \times S^7$ 
\be
\la{3.2}
ds^2 =\tfrac{1}{4}\, L^2 \big( dx^2 +  \sinh^2 x\, d\widetilde{S}_{2} +   \cosh^2 x\,  dy^2\big)
+   L^2 \big( dv^2 + \cos^2 v   \, dS_{5} + \sin^2 v \, dz^2 \big)  \ ,
\ee
where the boundary of  \tAdS$_4$     is $\widetilde S^2 \times S^{1}_{\beta} $  with  the $S^1_\b$ coordinate 
being $y\equiv y + \beta$   and 
\be
\la{3.3}
d\widetilde{S}_{2} = d\vartheta^{2}+\sin^{2}\vartheta\,(d\varphi+i 
dy)^{2}.
\ee
The  topologically non-trivial shift  of the $2\pi$-periodic angle $\varphi$  by $i y$  is required  by supersymmetry.\foot{As already mentioned in footnote \ref{f7}, 
this may be compared to the 11d metric AdS$_7 \times \widetilde{S}^4$ 
  in ref.  \cite{Beccaria:2023sph}   which discussed the M2  brane $ S^1_\beta\times S^2$ 
   instanton contribution to the   large $N$ expansion of the superconformal 
   index of   the  $(2,0)$  theory. 
  The roles of    $S^{2}$ and $S^{5}$  are thus effectively  
  interchanged.
Here the  twist is  in the AdS$_{4}$ part, while in   \cite{Beccaria:2023sph}
   it was
in the $S^{4}$  part  of AdS$_7 \times {S}^4$ (as it was 
 associated with the $R$-charge   generator in the definition of  the (2,0) index).}
The  metric \rf{3.2}   solves the 11d supergravity equations  when supplemented  by the "electric" 4-form flux 
\ba
\la{3.4}
F_{4} =& dC_{3} = -  \tfrac{3i}{8} L^3\, \vol_{_{\widetilde {\rm AdS}_{4}}}, \qquad 
\vol_{_{\widetilde {\rm AdS}_{4}}}= \sinh^{2}  x  \cosh x\,  d x \wedge dy\wedge \vol_{\widetilde{S}^{2}}= \vol_{_{{\rm AdS}_{4}}}, \notag \\
C_{3} =& -  \tfrac{i}{8}  L^3 \, \sinh^{3}x\,   dy \wedge \vol_{S^{2}},\qquad \ \ \ \ \ \ \  \  L^6=  32 \pi^2 N   \ell_{\rm P}^6,
\ea
where   the shift term  in $\vol_{\widetilde S^{2}} = \sin\vartheta\, d\vartheta \wedge (d\varphi+i dy)$ does not  contribute  to the product with $dy$. 

\subsection{M5 brane action}\la{sec:class}

The Euclidean action of a probe M5 brane in an 11d supergravity   background 
  may be written as  \cite{Bandos:1997ui,Aganagic:1997zq,Schwarz:1997mc} (see also \cite{Howe:1996yn,Bandos:1997ui,Claus:1998fh})
\ba
\la{3.6}
S =& T_{5}\, \int d^{6}\xi\,\Big[\sqrt{\det(G_{mn}+\widetilde{\rm H}_{mn})}-\tfrac{i}{4}\,\sqrt{G}\, \widetilde{\rm H}^{\star\, mn}\widetilde{\rm H}_{mn}\Big]
 +i\,T_{5}\int\Big(C_{6}+\tfrac{1}{2}{\rm H}\wedge C_{3}\Big)+ S_{\rm F} (\theta), \\
& H_{mn\ell} = 3\partial_{[m}A_{n\ell]}, \qquad {\rm H}_{mn\ell} = H_{mn\ell}-C_{3, mn\ell}, \qquad \ H^{\star\, mn\ell} = \frac{1}{6\,\sqrt{G}}\,\eps^{mn\ell abc }H_{abc},   \\
& \widetilde {\rm H}_{mn} = {\rm H}^{\star}_{mn\ell}\,U^{\ell}, \qquad \widetilde {\rm H}_{mn}^{\star} = {\rm H}_{mn\ell}U^{\ell}, 
\qquad U_{\ell}(\xi) \equiv   \frac{1}{\sqrt{(\partial_{r} a)^{2}}}\,\partial_{\ell}a(\xi)  \ .\no
\ea
Here 
 $G_{mn} = \del_m X^M \del_n X^N G_{MN} (X(\xi))$ is the induced metric  ($X^M$ are 11d coordinates), 
 $H_{mn\ell}$ (which is self-dual on shell) is the  field strength of the world-volume  antisymmetric gauge field $A_{mn}(\xi)$
 and $\theta$ denotes  the   11d  fermions ($S_{\rm F} (\theta)$  is given in  \rf{3.41}  below). The 6-form 
 $C_{6}$  is defined by\footnote{$F^\star _{4}$ is the 11d dual of $F_4$.
 Note also that $d(dC_{6})=0$ on the equations of motion  for  $C_{3}$  (assuming   there is no 11d  gravitino  
 background, see, e.g.,  \cite{Candiello:1993di}).}
\be
\la{3.5}
dC_{6} =  F^\star _{4}-\frac{1}{2} C_{3}\wedge F_{4} \ , 
\ee
The auxiliary scalar $a(\xi)$  ensures the manifest 6d covariance of the action. 
It   may  be fixed   by a gauge choice    $a(\xi)=\xi^{1}$ \cite{Pasti:1997gx}.


Below   we  will consider a  semiclassical expansion of  the path  integral  for  the  M5 brane 
 in the   \tAdS$_4\times S^7$   background \rf{3.2},\rf{3.4}
\ba
\la{x37}
&{\rm Z} =\int [dX\, dA_2\,  d\theta]\, e^{- S[X,A_2,\theta]} = \  \zZ_{1}\,  e^{- S_{\rm cl}}\big[1 + \mc O({ T}_5^{-1} )\big]   \ , \ \ \ \ \ \ \ \ \ \ \ 
S_{\rm cl} = { T}_5 \, \bar S_{\rm cl}  \ , \
\\
\la{x38}
&\zZ_{1} =  e^{-F} \ , \ \ \ \qquad \   F =\tfrac{1}{2}  \sum_r  (-1)^{{\rm F}_{r}} \log \det \Delta_r \ . 
\ea
Here $\Delta_r$   are   second-derivative  operators of quadratic fluctuations. 

The    action \rf{3.6}  admits  the  classical solution  corresponding to   the M5  brane wrapped on  $S^{1} \subset$   \tAdS$_5$\     and $ S^5 \subset S^{7}$  in \rf{3.2}  with other $X^M$  coordinates,
the $A_{mn}$  gauge field,  and the fermions being trivial 
\be \la{39}
X^1= y=\xi^1, \ \ \ \ \   X^i(S^5)= \xi^i\    (i=2,..., 6)  \ ,  \ \ \ \ \ \   x=0, \ \ \   v=0,\ \  \ \    A_{mn}=0, \ \ \  \theta=0 \ . 
\ee
The induced metric  (up to the overall $L^2$ factor) is then
\be \la{40}
G^{(0)}_{mn} d \xi^m d \xi^n =\tfrac{1}{ 4}  d \xi_1^2  + dS_5\, , \ \ \ \ \ \ \  \ \  dS_5 = g_{ij} (\xi) d \xi^i d \xi^j, \ \ \ \ \ \ 
\ \ \ \   \xi^1 \equiv \xi^1 + \beta  \ ,  \ee
where $dS_5$  is the metric  of a unit-radius 5-sphere. Note  that  
  the presence  of the $1\ov 4$   factor  in front of  $d \xi_1^2$  means that the effective  length of the  "thermal" circle is $\ha \beta$  rather than $\beta$. 

The corresponding classical   value of the M5 brane action may be written as 
\be
\la{3.14}
S_{\rm cl}  = S_{\rm V}+S_{\rm WZ} = T_{5}\,\int d^{6}\xi\, \sqrt{\det G}+i\,T_{5}\,\int C_{6} \ .
\ee
Here the 5-brane tension  enters together   with  the $L^6$  factor of the scale  in  \rf{3.2},\rf{3.4}\foot{Note that   the 5-brane tension is related to  the 2-brane tension $T_{2} = \frac{1}{(2\pi)^{2}\,\ell_{\rm P}^{3}}$   as \cite{Klebanov:1996mh}:   \ \ 
$T_{5} =   \frac{1}{2\pi}\,T_{2}^{2}$   and thus also 
${\rm T}_{5} =   \frac{1}{2\pi}\,{\rm T}_{2}^{2}$  where   ${\rm T}_{2} \equiv T_{2}L^{3}$. }
\be \la{314}
T_{5} = \frac{1}{(2\pi)^{5}\,\ell_{\rm P}^{6}} , \ \ \ \ \ \ \ \qquad 
{\rm T}_{5} \equiv  T_{5}L^{6} = \frac{1}{\pi^{3}} N\ .   
\ee
To find $C_6$ in \rf{3.5} 
 we note that \rf{3.4} implies that,  written in terms of vielbein   1-forms, the expression of  $F_4$ is\foot{Note that $ F^\star_{4}$ 
 is   defined 
 originally {in Minkowski space} where $F_{4}$ is real.  In the present case 
  $ F^\star_{4}$ is purely spatial and  thus remains real  after the  rotation to the Euclidean signature.}  
\be\la{317}
F_{4} = -6iL^3 \,e_{x} \wedge dy\wedge e_{\vartheta}\wedge e_{\varphi}\ , \ \ \ \ 
 F^\star_{4} =  -6L^6\,e_{v}\wedge e_{S^{5}}\wedge e_{z}  = -6L^6\,\cos^{5}v\,\sin v\, dv\wedge \vol_{S^{5}} \wedge dz,
\ee
where we used the explicit form of the $S^{7}$ metric  in (\ref{3.2}). As a result, 
\be
\la{3.22}
C_{6} =  L^{6}\,\cos^{6}v\,\vol_{S^{5}} \wedge dz\ .
\ee
The  WZ term in \rf{3.6},\rf{3.14}   then  vanishes on the classical solution \rf{39}  (but it will  contribute to   the  quadratic fluctuation  action). 
Using 
   that 
 $\vol(S^{5}) = \pi^{3}$  we thus get  
\be
\la{3.18}
S_{\rm cl}  =S_{\rm V}=   {\rm T_{5}} \,   \tfrac{1}{2}\beta\, \vol(S^{5}) = \tfrac{1}{2}\beta  N  \ . 
\ee
Thus  $e^{- S_{\rm cl}} = q^N$     where  $q= e^{ -\ha \beta}$. 
This   reproduces the leading  $q^N$ term in \rf{2.8} in agreement with  the  suggestion  \cite{Arai:2020uwd}
 that  the $S^1_\b \times S^5$  M5  instanton   should be 
  responsible for the leading   large $N$   correction to the  ABJM  index. 

Below  we will compute the one-loop correction 
\rf{x38}  to the M5 brane partition function expanded  near this 
classical solution. We will    choose the 
static gauge identifying  6  of the $X^M$  coordinates with the M5 world-volume coordinates 
 $\xi^m \ (m=1,\dots, 6)$, i.e. setting  to  zero the  fluctuations of $X^1$ and $X^i$ in \rf{39}. We   will  
   also fix  the   $\kappa$-symmetry gauge \rf{344}   for the fermions.
  We will   be left with  5 scalar fluctuation modes  of the transverse coordinates, 3  physical  modes  of $A_{mn}$ 
and 8 fermionic   modes.\foot{These physical degrees of freedom are 
the same as  of  a   6d $(2,0)$ tensor multiplet describing the 
leading 2-derivative term in the static-gauge M5 brane action  in flat space, \ie  five scalars, a real antisymmetric
 2-tensor with (anti)self-dual field strength, and four 
  symplectic Majorana--Weyl  6d spinors.
 They transform,  respectively,  in the $\bm{5}, \bm{1}, \bm{4}$ representations of  $USp(4)$  (see, e.g.,  \cite{Bergshoeff:1999db}).}
 
 The corresponding fluctuation  operators $\Delta = - \nabla^2 + ...$   in \rf{x38} will be defined   
 on $S^1_\b \times S^5$ with the metric \rf{40}  with periodic  $S^1_\b$ boundary conditions for fermions.\footnote{The discussion  of fluctuations for the bosons and fermions will be very similar to the one 
in the case of the   $S^1_\beta \times S^2$   M2 brane instanton contribution to the index
 of the $(2,0)$ theory in  \cite{Beccaria:2023sph}.}
 Their  contributions   to the one-loop  free energy \rf{x38}  may be  split  
 into the "Casimir"   part    linear in the 
 effective length $\ha\beta$ of the circle in \rf{40}
 and a   "thermal"   part  which is exponentially small at large $\beta$  (see,    e.g.,  \ci{Gibbons:2006ij}) 
\be
\la{3.10}
F(\beta) = \tfrac{1}{2}\b E_{c}+\bar F(\beta) \ .
\ee
 The term  $\bar F(\beta($ will have the 
form (here $k$ is   labelling eigenvalues  of a  fluctuation operator  on $S^5$)
\be
\la{3.11}
\bar F(\beta) = \sum_{k}c_{k}\log\big(1-e^{-\frac{1}{2}\b E_{k}}\big) =
-\sum_{m=1}^\infty {1\ov m} \,  \sum_{k} c_{k}\, e^{-\frac{1}{2}\b m E_{k}}\ . 
\ee
Thus  the one-loop partition function in \rf{x37},\rf{x38}    may be written as   
\be
\la{3.12}
\zZ_{1} = e^{-\frac{1}{2}\b E_{c}}\, \Z(q)\  ,  \qquad \qquad 
 \Z(q) = e^{-\bar F} =  \exp \big[ \sum_{m=1}^\infty {1\ov m} \Zsp(q^m)\big] \equiv \PE[\Zsp(q)],
\ee
where the  
 "single-particle" partition function $\Zsp(q)$   is   (\cf (\ref{2.4}))
\be
\la{3.13}
\Zsp(q) = \sum_{k}c_{k}\, q^{E_{k}}, \qquad \qquad \ \  q \equiv  e^{-\ha \beta}.
\ee
Each of the  one-loop  contribution of 6d  fluctuation  fields will  have no logarithmic  UV divergences with power divergences removed  by zeta-function regularization.    The 
 aim  will  be  to compare  the total  result for the partition function 
 with  the  prefactor of the  $q^N$ term in the
 ABJM  index in \rf{2.8},\rf{2.12},\rf{2.13}.


\def \rX  {{\rm X}}   \def \rY  {{\rm Y}}

\subsection{Scalar fluctuations}\la{sca}

Expanding the action \rf{3.6} to quadratic order in fluctuations around the  classical background \rf{39}  using  the static gauge we observe that as  the 3-form $C_3$  in  (\ref{3.4}) is  of higher order in fluctuation   fields its contribution to ${\rm H}_{mnk}$   as well as the second  part of the  WZ  term  may be ignored.  
The resulting scalar part of the M5 brane action  then  takes  the  form of  \rf{3.14}.

To recall, in the  static gauge   $y=\xi_{1} \in (0,\beta)$ and $\xi_{2}, \dots, \xi_{6}$ are coordinates on $S^{5}$. 
The 5 "transverse" angular   coordinate fields   $x(\xi), \theta(\xi), \varphi(\xi), v(\xi), z(\xi)$  can be 
expressed  in terms of 5 "Cartesian"  ones $\rX_1,\rX_2,\rX_3$ of  AdS$_4$ part  
 and $\rY_1,\rY_2$  of $S^7$   part defined as  
\ba
\rX_{1} =& \sinh x\,\cos\vartheta, \ \ \ \ \quad  
\rX_{2} = \sinh x\,\sin\vartheta\,\cos\varphi, \ \ \ \ \quad 
\rX_{3} = \sinh x\,\sin \vartheta\,\sin\varphi,\no  \\
& \qquad \rY_{1} = \sin v \cos z, \ \ \ \quad  \rY_{2 }= \sin v \sin z,\la{320}
\\
x =& \text{arcsinh}\,\sqrt{\rX_{1}^{2}+\rX_{2}^{2}+\rX_{3}^{2}},\qquad \vartheta = \arctan\frac{\sqrt{\rX_{2}^{2}+\rX_{3}^{2}}}{\rX_{1}}, \qquad \varphi = \arctan\frac{\rX_{3}}{\rX_{2}},\no  \\
v =& \arcsin\sqrt{\rY_{1}^{2}+\rY_{2}^{2}},\qquad \qquad  z = \arctan\frac{\rY_{2}}{\rY_{1}}.\la{321}
\ea
For generality, let us consider the case when  the  shift of $\varphi$  by $i   y$ has  a  coefficient $\kappa$ 
(we will   set $\kappa$ to its "supersymmetric"  value  $\kappa=1$ at the end). 
To expand the  volume  $S_{\rm V}$  part of   \rf{3.6},(\ref{3.14}), we   note that according to \rf{321} 
\ba
& \tfrac{1}{4}\, \Big[ dx^2 +  \sinh^2 x\, \big(d\vartheta^{2}+\sin^{2}\vartheta\,(d\varphi+i \kappa\, d\xi_{1})^{2}\big) +   \cosh^2 x\,  d\xi_{1}^2\Big] \lp
= \tfrac{1}{4}d\xi_{1}^{2}\,[1+\rX_{1}^{2}+(1-\kappa^{2})(\rX_{2}^{2}+\rX_{3}^{2})]+\tfrac{1}{4}d\rX_{a}^{2}+\tfrac{i}{2}\kappa\,d\xi_{1}\,(\rX_{2}d\rX_{3}-\rX_{3}d\rX_{2})+\cdots,\la{322}
\\
\la{3.28}
&dv^2 + \cos^2 v   \, dS_{5} + \sin^2 v \, dz^2 = dS_{5}\,(1-\rY_{a}^{2})+d\rY_{a}^{2}+\cdots.
\ea
The perturbed induced metric is then 
\be
G_{mn}\,d\xi^{m}d\xi^{n} = (G_{mn}^{(0)}+G_{mn}^{(1)}+\cdots)\,d\xi^{m}\,d\xi^{n}, \qquad m,n=1,\dots, 6 \ , 
\ee
where  $G_{mn}^{(0)}$   is the   metric of $S^1\times S^5$   given in \rf{40}.
We shall use  $g_{ij}\equiv g_{ij}({S^{5}})$  ($i,j=2, ...,6$)   and  denote the derivative over the $S^1$ direction 
 $\xi^1$ as  $\partial_{1}\Phi \equiv \dot\Phi$. 
 After an  integration by parts and rescaling $\rX_p$  ($p=1,2,3$)   and $\rY_r$  ($r=1,2$) 
   fields  by constant factors (including $({\rm T}_5)^{-1/2}$) 
   we get   for the corresponding  quadratic  fluctuation contributions to 
   $S_{\rm  V}$ 
\ba
&\delta S_{\rm V} = \delta S_{\rm V}({\rX})+\delta S_{\rm V}({\rY}),  \no \\
\la{3.31}
\delta S_{\rm V}({\rX}) =& \int d^{6}\xi\,\sqrt{g}\,\Big(
\tfrac{1}{2}g^{ij}\partial_{i}\rX_{p}\partial_{j}\rX_{p}
+2\rX_{1}^{2}+2(1-\kappa^{2})(\rX_{2}^{2}+\rX_{3}^{2})+2\dot \rX_{a}^{2}+8i\kappa\,\rX_{2}\dot \rX_{3} 
\Big), \\
\delta S_{\rm V}({\rY}) =& \int d^{6}\xi\,\sqrt{g}\,\Big(
\tfrac{1}{2}g^{ij}\partial_{i}\rY_{r}\partial_{j}\rY_{r}
+2\,\dot \rY_{r}^{2} -\tfrac{5}{2}\,\rY_{r}^{2}
\Big)\ . \la{325}
\ea
To find the  contribution of the  WZ term in \rf{3.14}  we  use \rf{3.22}  
(integrating by parts   and   again doing constant field rescaling)
\be
\la{3.33}
i T_5 \int  C_{6} \to  \int \,\vol_{S^{5}}\wedge [3\times 2\times (\rY_{2}d\rY_{1}-\rY_{1}d\rY_{2}) ]= \int 12\, \,\rY_{2}\dot \rY_{1}\, \vol_{S^{5}}\wedge d\xi_{1}\ . 
\ee
Combining this  with \rf{325}   we get 
\ba
\la{3.34}
\delta S_{\rm V}({\rY})  + S_{\rm WZ}({\rY}) =& \int d^{6}\xi\,\sqrt{g}\,\Big(
\tfrac{1}{2}g^{ij}\partial_{i}\rY_{r}\partial_{j}\rY_{r}
+2\,\dot \rY_{r}^{2} -\tfrac{5}{2}\,\rY_{r}^{2}+12i\, \rY_{2}\dot \rY_{1}
\Big).
\ea
Next, let us   expand the fields $\rX_p$ and $\rY_r$   in Fourier series in $S^1$, i.e. 
$\rX_1=\sum_{n=-\infty}^\infty \rX^{(n)}_1 \, e^{i  \nb\xi_1}$,  etc.,  so that 
 $\partial_{1}\to i\nb$, where 
\be\la{328} 
\nb \equiv  \frac{2\pi n}{\beta}, \qquad n\in\mathbb Z. 
\ee
As a result, we get towers of  scalar fields   on $S^5$   with the   standard kinetic operators $-\nabla^2$ on $S^5$ and the 
effective mass terms depending on $\nb$.  Explicitly, from \rf{3.31} 
  $\rX_{1}$  then  leads to  a  scalar  tower on $S^{5}$ with the  mass squared
\be
M_{\rX_{1}}^{2} = 4+4\,\nb^{2} \ . \la{329}
\ee
Since the   6d  Lagrangian for  $\rX_{2}, \rX_{3}$  in \rf{3.31}  may be written as  
\be
\mathscr L_{\rX} = g^{ij}\partial_{i}\bar\rX\partial_{j}\rX- 4\bar\rX\,\big[\ddot \rX+2\kappa\dot\rX+(\kappa^{2}-1)\rX\big]\  ,
\qquad \rX =  \frac{1}{\sqrt 2} (\rX_{2}+i\rX_{3}) \ ,  \ee
the corresponding   mass   in the action on $S^5$ will be  given by (we now specialize to the case of $\kappa=1$)
\be
\la{3.38}
M_{\rX}^{2} =  4+4\,(\nb+i\kappa)^{2} \Big|_{\kappa=1} = 4+4\,(\nb+i)^{2} \ . 
\ee
Similarly,  from (\ref{3.34}) for $\rY_r$  we get 
\be
\la{3.39}
\mathscr L_{\rY} = g^{ij}\partial_{i}\bar\rY\partial_{j}\rY+\bar\rY\,(-5\rY-4\ddot\rY-12 \dot\rY)\ ,
\qquad   \rY= \frac{1}{\sqrt 2}(\rY_1 + i \rY_2) \ ,  \ee
corresponding to  
\be
\la{3.40}
M_{\rY}^{2} = -5+4\nb^{2}-12i\,\nb = 4+4(\nb-\tfrac{3}{2}i)^{2}\ .
\ee
In \rf{329},\rf{3.38},\rf{3.40}  we isolated 
the  $M^2 =4$  term that corresponds to  a  conformally coupled  massless scalar  in 6d space 
$S^1 \times  S^5$  (cf. footnote \ref{foot:lapl} below). 
The factor 4  that multiplies the $\nb^2$ terms in the masses in \rf{329},\rf{3.38},\rf{3.40} 
originates  from the $1\ov 4$ in the   induced metric \rf{40}\foot{Explicitly, we have $-\nabla^{2}_{S_{1}\times S_{5}} = -4 \partial_{1}^{2}-\nabla^{2}_{S^{5}}$.}
  and  corresponds to 
the effective  rescaling  $\beta \to \ha \beta$  already  mentioned above.

Note  also that  while  the  shift of $\nb$  in (\ref{3.38}) is a  direct consequence of the  twist 
of the $S^{2}$  angle in (\ref{3.3}) 
  implying  $\xi^1$-dependent rotation in the $(\rX_2, \rX_3)$  plane, 
    the non-trivial   shift  in (\ref{3.40}) is due to the contribution of the  WZ term (\ref{3.33}).
The two are  effectively correlated  as a result of the 
 underlying supersymmetry of the problem.\foot{Note that the signs of the 
shifts in \rf{3.38}  and \rf{3.40}  do not matter as $\nb$ takes both   positive and negative values.}

\subsection{Fermionic  action}\la{fer}

The   quadratic fermionic part  of the  $\kappa$-symmetric M5  brane  action
in a general  background  which is solution of 11d  supergravity 
may be written as   \ci{Bandos:1997ui,Aganagic:1997zq}  (here $m,n=1, ...,6$   as in \rf{3.6},  $X^M$ are 11d coordinates
and $\theta$  is a 32 component 11d Majorana spinor) 
    \ba
 & S_F =i T_5  \int d^6\xi\Big[ \sqrt G \,   G^{mn}   \del_m X^M  \,   \bar \theta \,  \Gamma_M \hat D_n \theta \nonumber\\
&\qquad \qquad \qquad\ \ \   - \tfrac{1}{ 5!}  \eps^{mnklpq}  \del_m X^M  \del_n X^N \del_k X^K  \del_l X^L  \del_p X^R   \,  \bar \theta \,  
 \Gamma_{MNKLR}  \hat D_q \theta   + {\cal O}(\theta^4) \Big]
  \ , \la{3.41} \\
  & G_{mn} = \del_m X^M  \del_n X^N   G_{MN} (X), \qquad \ \  \
   G_{MN} = E^A_M E^A_N, \qquad  \ \ \
  \Gamma_M  =   E_M^A(X) \Gamma_A\ ,    \la{3.42}
  \\
  &   \hat D_k = \del_k X^M   \hat D_M , \ \qquad 
  \hat D_M = {\rm D}_M  - \tfrac{1}{288} (\G^{PNKL}_{\ \ \ \ \ \   \ M} + 8  \G^{PNK}\delta^L_M ) F_{PNKL} \ .  \la{3.43}
\ea
  $ \hat D_M $  is the   generalized 11d  spinor covariant derivative  \ci{Cremmer:1979up} 
and  ${\rm D}_M=\del_M + \four \G_{AB} \omega^{AB}_M $.
In particular  cases of the   maximally supersymmetric AdS$_4 \times S^7$
 or   AdS$_7 \times S^4$ backgrounds   the  fermionic part of  the M5 brane action may be written in an explicit form 
including also   higher orders in $\theta$ \ci{Claus:1997cq,deWit:1998yu,Claus:1998fh,Sakaguchi:2004bu}.

In the present case of the M5   brane wrapped on $S^1_\beta \times S^5$   in the "twisted"  \tAdS$_4\times S^7$ background \rf{3.2},
the computation  of the fermionic contribution to the one-loop  effective action is very similar 
to  the one in the case of the     M2  brane instanton wrapped on $S^1_\beta \times S^2 $  in AdS$_7 \times {\widetilde S}^4$  
considered in  \ci{Beccaria:2023sph}.  As  discussed above, we fix 
 the static gauge  $X^m= \xi^m, \ X^i=0$  ($i=2, ...,6$).  
A  natural  choice for a  $\kappa$-symmetry   gauge is like  in the flat  space case, i.e.   
\be \la{344}
(1- \G) \theta=0\ , \qquad \qquad  
\G\equiv {1\ov 6! \sqrt G}  \eps^{mnklpq}  \del_m X^M  \del_n X^N \del_k X^K  \del_l X^L  
\del_p X^R  \del_q X^P \Gamma_{MNKLRP}\ . \ee
Then the WZ term in \rf{3.41} takes the same form as the  Dirac term 
with the number of independent  components of $\theta$ reduced to 16.

Like in  the M2 brane case in \ci{Beccaria:2023sph} the resulting fermionic action 
 may be represented as a     collection of  4 MW 
6d   fermions (each with 4 real components) 
on  $S^1 \times S^5$  with  the  kinetic   operator  
$i \G^k {\nabla}_k + \Mp  $
coming from  $\hat D_n$ in \rf{3.41}. 
 The  "mass"  operator $ \Mp$    contains  two 
 operators $P_1,P_2$ built out of $\Gamma$-matrices that square to one and thus can be used 
 to define projectors.
 One 
  originates from  the  contribution to \rf{3.43}  of the $F_4$ flux along the AdS$_4$  directions  in \rf{3.4} 
 (i.e. it  is  proportional to $\Gamma_{x\vartheta y \varphi}$)
 and the other is 
 related to the contribution  of the  constant term in the 
 Lorentz connection related to the rotation in \rf{3.3} (i.e. it is  proportional   to $\Gamma_{y \varphi}$)
 \be \Mp =    \tfrac{3}{ 2} P_1 + P_2\ , \qquad  P_1^2=P_2^2=1\ , \qquad  [P_1, P_2]=0 \ . \la{3550}
 \ee 
 Diagonalizing $\Mp$ we get   four  6d   MW  fermions   with the  effective masses 
 $ \pm {3\ov 2} \pm 1$, i.e. $\pm {1\ov 2}$ and $ \pm  {5\ov 2}$. 
 
 Recalling  that  the vielbein  component  corresponding to 
 the $y$ direction in the  induced 6d metric  originating from \rf{3.2}  has an extra $\ha$  relative to the $S^5$ part 
 (so that inverse vielbein  comes with a  factor of 2)  we get 
 $\G^m {\del}_m \to  2  \Gamma^1 \del_1   + \Gamma^j \del_j$ 
 where $j$   is the   index of the $S^5$ directions. 
 Expanding  the 6d fermions  in Fourier modes in $\xi^1$   implies that  $\partial_1 \to  i \nb $ 
 (cf. \rf{328}) 
 and we thus end up with four  towers of  the 5d  fermionic  fields    on $S^5$ 
 with  the operators 
 \be  {\cal D}=   i \G^j {\nabla}_j + M_{\psi} , \ \ \ \ \  \ \ \ \   
 M_{\psi} = 2 ( \nb + i\nu), \ \ \ \ \ \ \ \ \ \ 
 \nu =  \te \pm  {1\ov 4}, \ \  \pm   {5\ov 4}\ .\la{3.44}\ee
 Since the $S^1_\beta$ mode number $\nb$  takes both positive and negative values
 we may  assume that $\nu$ takes   just positive values ${1\ov 4}$ and $ {5\ov 4}$. 
 The corresponding   squared fermionic operator  has the form 
  \be \la{3444}
  \Delta_{1\ov 2} = - {\hat \nabla}_{S^5}^2 + \tfrac{1}{4} R   + M^2_{\psi} =  - {\hat \nabla}_{S^5}^2 + 5    + M^2_{\psi}\ , 
  \ee 
       where  $\hat \nabla$ contains spinor connection   and 
       we used that  the scalar curvature is $R(S^1\times S^5)=R(S^5) = 20$.

\subsection{One-loop free energies} 
\la{sec:free-sc}

Having found   the  scalar and fermion   quadratic fluctuation operators on $S^1  \times S^5$ 
 we may now compute the corresponding 
 contributions to the one-loop free energy in \rf{x38},\rf{3.12},\rf{3.13}. 
 We will  then  include also the   contribution of the antisymmetric tensor field  
 which is straightforward to find as it  does not involve a twist or coupling to $C_3$.


\subsubsection{Scalars}

The 5 scalar fluctuations written in terms of the Fourier   modes  on $S^1$ as towers of fields on $S^5$  have effective   masses
(see \rf{329},\rf{3.38},\rf{3.40}) 
\ba
\la{3.45}
&M^2 = 4 + \bar M_n^2, \ \ \ \ \ \  \bar M^{2}_n =  4\,(\nb+i\nu)^{2},\qquad  \nb=\frac{2\pi}{\beta}\,n,\ \ n\in\mathbb Z, 
\\ \la{3.46}
& \nu(\rX_{1}) = 0, \qquad \nu(\rX_{2})=\nu(\rX_{3}) = 1, \qquad \nu(\rY_{1})=\nu(\rY_{2}) = \tfrac{3}{2} \ .
\ea
The   contribution 
 to  free energy  from each scalar  fluctuation field    is then\footnote{\la{foot:lapl}We use  label $\phi$ to indicate the scalar   contribution.
In general, the operator $-\nabla^{2}_{S^{d}}$  on a  unit-radius $d$-sphere 
 has eigenvalues $\lambda_{k}$ with degeneracy ${\rm d}_{k}$ (see \cite{Rubin:1984,Higuchi:1986wu,Camporesi:1994ga})
\be\notag
\lambda_{k}=k(k+d-1), \qquad {\rm d}_{k} = \frac{(2k+d-1)\,(k+d-2)!}{k!\,(d-1)!}, \qquad k=0,1,2,\dots\, .
\ee
Thus  for a conformally coupled  (in $S^1 \times S^d$) 
 scalar  operator  
$-\nabla^{2}_{S^{d}}  + {1\ov 4}  (d-1)^2 $  one gets   $\lambda_k \to  
\lambda_{k}+\frac{1}{4}(d-1)^{2} = (k+\frac{d-1}{2})^{2}$  or $(k+2)^2$   in the present case of $d=5$.}
%
\ba
\la{3.48}
F^{S^{1}\times S^{5}}_\phi (\beta,\nu) =&\sum_{n\in \mathbb Z} F^{S^5}_\phi (\b, \bar M_{n}) = \tfrac{1}{2} \sum_{n\in \mathbb Z}  \log \det ( - \nabla^2_{S^5} + 4 + \bar M^2_n) \no \\
=& \tfrac{1}{24}\sum_{n\in\mathbb Z}\sum_{k=0}^{\infty}(k+1)(k+2)^{2}(k+3)\,\log\big[(k+2)^{2}+4(\nb+i\nu)^{2}\big] \ . 
\ea
The double sum in (\ref{3.48}) is formally  divergent  and can be defined   using the standard $\zeta$-function 
 regularization (see Appendix \ref{app:free-energy}).  As a result, we get (cf.   (\ref{3.10}))
\ba
\la{3.49}
&F^{S^{1}\times S^{5}}_\phi(\beta, \nu) = \tfrac{1}{2}\b \,E_{c,\phi}(\nu) + \bar{F}^{S^{1}\times S^{5}}_\phi(\beta), 
\\ 
\la{3.50}
& \bar{F}^{S^{1}\times S^{5}}_\phi(\beta, \nu) = \tfrac{1}{24}\sum_{\pm} \sum_{k=0}^{\infty}(k+1)(k+2)^{2}(k+3)\,\log\big[1-e^{-(k+2\pm 2\nu)\frac{\beta}{2}}\big]\ , \\
&E_{c,\phi}(\nu) = -\tfrac{31}{60480}+\tfrac{1}{18}\nu^{4}-\tfrac{4}{45} \nu^{6}\ . \la{351}
\ea
From $\bar F$, we read off  the  corresponding single-particle partition function  (see  (\ref{3.11}), (\ref{3.13}))
\be
\la{3.51}
\Zsp_{\phi}(q,\nu) = \tfrac{1}{24}\sum_{\pm}\sum_{k=0}^{\infty}(k+1)(k+2)^{2}(k+3)\, q^{k+2\pm 2\nu}, \qquad \ \ \ \ q = e^{-\ha \beta}.
\ee
Computing the sum, we obtain\footnote{Note that in 
\la{KL}
the conformal coupling case, i.e. 
  for $\nu=0$, the expression (\ref{3.51}),\rf{3.46}, i.e.
      $\Zsp_{\phi}(q,0) = \frac{q^{2}(1-q^2)}{(1-q)^{6}} = \frac{q^{2}(1+q)}{(1-q)^{5}}$
    can be obtained by counting  scalar  operators on $\mathbb{R}^{6}$, see Appendix E.} 
\be\la{346}
\Zsp_{\phi}(q,\nu) = \frac{q^{2}(1+q)}{2(1-q)^{5}}\,(q^{2\nu}+q^{-2\nu}).
\ee
The total contribution of the five scalars   corresponding to \rf{3.46} 
is then 
\be\la{347}
\Zsp_{\rm scalar}(q) = \Zsp_{\phi}(q,0)+2\Zsp_{\phi}(q,1)+2\Zsp_{\phi}(q,\tfrac{3}{2})  = \frac{(1+q)(1+q+q^{3}+q^{5}+q^{6})}{q\,(1-q)^{5}}.
\ee
The sum of  the scalar Casimir energies  is  given by 
\be
E_{c, \rm scalar} = E_{c, \phi}(0)+2E_{c, \phi}(1)+2E_{c,\phi}(\tfrac{3}{2}) =  -\tfrac{92639}{60480}\ .\la{349}
\ee

\subsubsection{Fermions}

The discussion of the fermionic  contribution to free energy  corresponding to  \rf{3.44},\rf{3444}  is similar. 
We  assume that fermions are periodic on $S^1$ as  appropriate for   a 
 "supersymmetric"  partition function. 

Let us consider first  a conformally coupled, i.e. massless,    spinor  field on $S^{1}\times S^{5}$. 
Using that the eigenvalues and degeneracies of the operator  $-\hat \nabla^{2}_{S^{d}}$  with spinor covariant  derivative
are  given by 
 \cite{Camporesi:1995fb}
\be\la{3515}
\l_{k} = \Big(k+\frac{d}{2}\Big)^{2}-\frac{d(d-1)}{4}, \qquad\qquad  {\rm d}_{k} = \frac{(k+d-1)!}{k!(d-1)!},\qquad k=0,1,2,\dots\, , 
\ee
the eigenvalues of $-\hnabla^{2}_{S^{d}}+\frac{1}{4}R = -\hnabla^{2}_{S^{d}}+\frac{d(d-1)}{4}$ are simply $(k+\frac{d}{2})^{2}$ (with the same degeneracy).
%
%
For $d=5$ this leads to  the  expression for the free energy  contribution of a single MW spinor 
  analogous to the   one in the scalar case  (\ref{3.49})\foot{Here the factor 4  corresponds to  4 real  components of a MW   spinor. Note that  for large $k$ 
the summand here is the same as 4 times the one  in real scalar  case  in \rf{3.48}.} 
 \ba
\la{3.59}
\bar F^{S^{1}\times S^{5}}_\psi(\beta, \nu ) =&
4\times \tfrac{1}{24}\sum_{n\in\mathbb Z}\sum_{k=0}^{\infty}(k+1)(k+2)(k+3)(k+4)\,\log\big[(k+\tfrac{5}{2})^{2}+4(\nb+i\nu)^{2}\big]\ . 
\ea
The sum over $n$ here  is computed as in the scalar case  and  one finds 
\be\la{356}
\bar F^{S^{1}\times S^{5}}_\psi(\beta, \nu) = \tfrac{1}{12}\sum_{\pm}\sum_{k=0}^{\infty}(k+1)(k+2)(k+3)(k+4) \log\big[1-e^{-(k+\frac{5}{2}\pm 2\nu)\frac{\beta}{2}}\big]\ .
\ee
The  corresponding single-particle partition function is\foot{For $\nu=0$ 
 the  single-particle partition function of a massless 6d fermion  on $S^1\times S^5$  can be found
 by conformal mapping to $\mathbb{R}^{6}$  and operator counting   giving 
$\Zsp_{\psi}(q,0) = \frac{4q^{{5}/{2}} (1-q)}{(1-q)^{6}}= \frac{4q^{{5}/{2}}}{(1-q)^{5}},
$
 as in  Eq.~(A.4) in \cite{Beccaria:2014qea}.
           }  
\be\la{354} 
\Zsp_{\psi}(q,\nu) = \frac{2q^{\frac{5}{2}}}{(1-q)^{5}}\,\big(q^{2\nu}+q^{-2\nu}\big)\ .
\ee
The Casimir energy  can be found   as in   (\ref{dd8})  and   $E^+_{c}(\nu)= E^-_{c}(\nu)$  so that 
\ba
E_{c,\psi}(\nu)   =& \frac{1}{12}\sum_{k=0}^{\infty}(k+1)(k+2)(k+3)(k+4)(k+\tfrac{5}{2}+2\nu)^{s}\ \ 
\stackrel{s\to 1}{\to}\ \   \tfrac{367}{96768}-\tfrac{3}{32}\nu^{2}+\tfrac{5}{18}\nu^{4} -\tfrac{8}{45}\nu^{6}\ . \la{355}
\ea
According to \rf{3.44}  the total contribution of two  MW spinors   with $\nu={1\ov 4} $ and two  with  $\nu={5\ov 4}$ 
  is then given by (we reverse the sign to account for fermion statistics)
\ba
& -2\,\Big[\Zsp_{\psi}(q, \tfrac{1}{4}) +\Zsp_{\psi}(q, \tfrac{5}{4})\Big]  
= - {4 (1 + q^2 + q^3 + q^5)\ov (1-q)^5 } \  \ , \la{3568} \ \qquad  -2\,\Big[ E_{c,\psi}( \tfrac{1}{4}) +E_{c,\psi}( \tfrac{5}{4})\Big]  = \te {2173\ov 7560} \ . 
\ea

\subsubsection{Antisymmetric tensor}

The  contribution of  the $A_{mn}$  gauge field  in \rf{3.6}  to the one-loop partition function on $S^1 \times S^5$ 
is straightforward to find  by first relaxing the self-duality   condition on  its strength $H_{mnk}$ and 
 taking  half of the resulting value  of free energy at the end (\cf     \cite{Bastianelli:2000hi}
and references there). Rather than computing  the    relevant determinants directly 
one may use the underlying conformal 
symmetry  (and the absence of conformal anomaly  in the  $S^{1} \times  S^{5}$ case)   to  map the 
problem to  scaling dimension  counting on $\mathbb{R}^6$  (\cf footnote \ref{KL})
and then  use the expression   for the   corresponding partition function in \ci{Kutasov:2000td,Beccaria:2014qea}. 

As a result,   the single-particle partition function 
for the self-dual  antisymmetric tensor on $S^{1}_{\beta}\times S^{5}$  is  found to be\foot{Here again
$q=e^{-\ha \beta}$ as in (\ref{3.51}),  taking into account that the  radii of 
 $S^{1}$  and $S^{5}$  differ by 1/2.}
\be
\la{3.55}
\Zsp_{A}(q) = \tfrac{1}{4}\sum_{k=0}^{\infty}(k+1)(k+2)(k+4)(k+5)q^{k+3} = \frac{10q^{3}-15q^{4}+6q^{5}-q^{6}}{(1-q)^{6}} = \frac{q^{3}(10-5q+q^{2})}{(1-q)^{5}} \ . 
\ee
The  corresponding   contribution to the  free energy is\foot{Note that  the large $k$ limit of the summand here is 3 times the one  for a real scalar in \rf{3.50}, in agreement with the fact that  a  2-form field with self-dual  field strength has 3  physical degrees of freedom.}  
\be
\la{3.56}
\bar F_{A}^{S^{1}\times S^{5}}(\beta) = -\sum_{m=1}^{\infty}\frac{1}{m} \Zsp_{A}(q^{m}) =  \tfrac{1}{4}\sum_{k=0}^{\infty}(k+1)(k+2)(k+4)(k+5)\log(1-q^{k+3})\ .
\ee
The Casimir energy is \ci{Beccaria:2014qea}
\be\la{360}
E_{c, A} = -\tfrac{191}{4032}.  
\ee

\subsubsection{Summary}

To summarize, we collect in Table \ref{t1} the expressions for  the single-particle partition functions
and   Casimir energies for all of  the above  fields.\foot{To recall,  here the fermions are  assumed to be  periodic   on $S^1_\b$ 
so   these are   contributions to a supersymmetric partition function and supersymmetric Casimir energy.
Note that ref.  \cite{Beccaria:2014qea}  considered instead 
the  standard thermal partition function of a (2,0) multiplet   with antiperiodic fermions.}
\begin{table}[H]
\be
\def\arraystretch{1.3}
\begin{array}{cccc}
\toprule
 & \Zsp(q, \nu) &  \nu & E_{c} \\
\midrule
\text{scalar} & \frac{q^{2}(1+q)}{2(1-q)^{5}}\,(q^{2\nu}+q^{-2\nu})  & 
0, \ 2\times 1, \ 2\times  \frac{3}{2}
& -\frac{31}{60480}+\frac{1}{18}\nu^{4}-\frac{4}{45}\nu^{6} \\
\midrule
\text{self dual tensor} & \frac{q^{3}(10-5q+q^{2})}{(1-q)^{5}} & 0 & -\frac{191}{4032} \\
\midrule
\text{fermion} &  \frac{2q^{\frac{5}{2}}}{(1-q)^{5}}\,(q^{2\nu}+q^{-2\nu}) & 2\times \frac{1}{4}, \ 2\times \frac{5}{4}
& \frac{367}{96768}-\frac{3\nu^{2}}{32}+\frac{5}{18}\nu^{4}-\frac{8}{45}\nu^{6}\\
\bottomrule
\end{array}\notag
\ee
 \caption{Single-particle partition function, shift $\nu$ and Casimir energy for  the fluctuation fields. \la{t1}
}
\end{table}
These   are to  be   combined  taking into account that fermions  contribute with opposite sign, i.e.
\be
\Zsp_{\rm tot} = \sum(-1)^{\rm F}\,\Zsp\ , \qquad\qquad 
E_{c, \rm tot} = \sum(-1)^{\rm F}\,E_{c}\ .\la{361}
\ee

\section{Partition function of   M5 brane  instanton and   ABJM  index}

As we already  saw in \rf{3.18}, the classical action    contribution $e^{-S_{\rm cl}}$
  of the $S^1 \times S^5$   M5 brane   
reproduces   the   factor $q^N$  in the ABJM index \rf{2.8}.
Let us now   present  the  total   result  for the  corresponding 
 one-loop partition function  
 by combining the contributions of the individual fluctuations    found  in the previous section.
  We will then  compare it with the prefactor of the 
 $q^N$  term 
   in  the ABJM index  \rf{1.2},\rf{2.12}    which is the $\bm{u}=\bm{1}$  limit of the  
   $q^N \delta\I_{N}^{(1)} $  part of (\ref{2.8}). 


\subsection{One-loop partition function  including  the zero-mode  contribution}   

Using the data in Table \ref{t1}  or combining together 
\rf{347},\rf{356},\rf{355} and \rf{349},\rf{3568},\rf{360} 
 we get   for the one-loop free energy  \rf{x38} (see  \rf{3.10}--\rf{3.13}) 
\ba
\la{x41}
F =& \tfrac{1}{2}\b E_{c}+\bar F, \qquad  \qquad 
\bar F = -\sum_{m=1}^{\infty}\frac{1}{m}\Zsp(q^{m}),\\ 
\la{4.1}
\Zsp(q) =& \Zsp_{\phi}(q,0)+2\Zsp_{\phi}(q, 1)+2\Zsp_{\phi}(q, \tfrac{3}{2})+\Zsp_{A}(q)
-2\,\Big[\Zsp_{\psi}(q, \tfrac{1}{4}) +\Zsp_{\psi}(q, \tfrac{5}{4})\Big] 
= \frac{7q}{1-q}+3+q^{-1}-q, \\
\la{4.2}
E_{c} =& E_{c,\phi}(0)+2E_{c,\phi}(1)+2E_{c,\phi}(\tfrac{3}{2})+E_{c, A}-2\Big[ E_{c,\psi}(\tfrac{1}{4})+E_{c,\psi}(\tfrac{5}{4})\Big] = -\tfrac{31}{24}.
\ea
The presence of  the constant   term 3  in \rf{4.1}   implies that the sum in the   expression for $\bar F$ in \rf{x41}
is formally divergent. 
This divergence can be attributed  to the  contribution of the zero modes of the fluctuation 
operators   that were   not   separated  in our formal treatment   of the "thermal" part $\bar F$ of the free  energy 
   in the previous section.  
   The  degeneracy   implying the presence of these  zero modes can be "resolved"   by switching on extra twists in the background (corresponding to non-trivial values of fugacities in the definition of the index)
  leading to a  finite  expression  for $\bar F$  (see section 4.3  below). 

As follows from the expressions   for the    fluctuation determinants  
in \rf{3.48},\rf{3.59},\rf{3.56},
%
the    zero modes appear  only  for the  scalars with $\nu= 1, \ {3\ov 2}$   and fermions   with $\nu ={5\ov 4}$
 and  only in the  sector with $\nb=0$ (i.e. on  the  "ground" level  of the $S^5$   KK modes).
  The   corresponding values of the 
$S^5$  mode  number $k$ and  their multiplicities   are summarized in Table \ref{t2}. 
\begin{table}[H]
\be
\def\arraystretch{1.3}
\begin{array}{cccc}
\toprule
\text{field} & \nu & k & \text{no.}\times {\rm d}_{k}  \\
\midrule
2\times\phi & 1 & 0 & 2\times 1 \\
2\times\phi & \frac{3}{2} & 1 & 2\times 6 \\
\midrule
2\times \psi & \frac{5}{4} & 0 & 2\times 4 \\
\bottomrule
\end{array}\notag
\ee
\caption{Scalar and fermion zero modes and their multiplicity. \la{t2} }
\end{table}
We thus get 14 bosonic  and 8 fermionic  zero modes  that should have  been 
 separated in the one-loop determinants.\foot{As  usual, 
their presence reflects (super)symmetries   preserved   by  the  corresponding  $S^1 \times S^5$   solution  embedded into AdS$4 \times S^7$   in \rf{3.2}.} 
 
Assuming that  the resulting $0^8/0^8$   ambiguity   can be  resolved,  the effective number of the remaining 
 (bosonic) zero modes is   6.
   This number   matches  the  value 3 of the constant term  in \rf{4.1}. 
   Indeed, regularizing the $k=0$  zero-mode term   in the scalar   case 
  in \rf{3.50}  by setting $\nu=1- \eps, \ \eps \to 0$  we get 
  $\bar F_0 = {1\ov 2} \log ( 1 - e^{-\eps\beta}) 
  = {1\ov 2} \sum_{m=1}^\infty {1\ov m}   e^{-\eps \b m} $. 
  Thus each zero mode contributes $1\ov 2$ to a   constant term in $\Zsp$.
  
 Let us 
note   that, as follows  from \rf{3.48},   for  $\nu= {3\ov 2} $  we get also  one   negative mode 
 for each of the two    $Y$-scalars  in  \rf{3.39},\rf{347},\rf{4.1}
 representing the transverse  $S^7$ fluctuations of M5 brane wrapped on $S^5\subset S^7$.
 This  reflects   the expected  instability of this M5 brane configuration. 
 Analytically continued, each of these negative modes   produces  an imaginary contribution to the resulting partition function. 
 Their  contribution  corresponds to  the $q^{-1}$ term in the single-particle partition function \rf{346}  (or \rf{347}
 and thus \rf{4.1}). It  translates   into $\sum_{m=1} {1\ov m} q^{-m}  =  - \log ( 1 - q^{-1}) $ term 
 in  the free energy   which is then  indeed
  imaginary for $q<1$. This issue   may be  formally ignored by  assuming   an analytic continuation in $q$
  as is usually done when computing the index.

With   the  standard  normalization of the   M5  brane path integral \rf{x37},   each zero mode 
is expected to contribute a factor  of $ \sqrt{{\rm T}_{5}} \sim \sqrt N $  (cf. \rf{3.6},\rf{314}). 
This  suggests that the resulting  expression for  the one-loop partition  function in \rf{x37} 
may be written as 
\ba
\la{x477}
&{\rm Z}_1=  Z_1  e^{-S_{\rm cl}}  = c_{0}\,N^{3}\,q^{N}\, q^{-{31\ov 24}} \,   e^{-\bar F'} \ , \\
&
\bar F' = -\sum_{n=1}^{\infty}\frac{1}{m}\big[\Zsp(q^{m})-3\big] = -\log(-q)-7\sum_{m=1}^{\infty}\frac{1}{m}\frac{q^{m}}{1-q^{m}}
\la{45}\ . 
\ea
In \rf{x477}   we explicitly included   the   zero-mode,   classical, and  the 
Casimir energy $q^{E_c}$  \rf{4.2}  contributions   and in \rf{45}  we subtracted  the  constant term in \rf{4.1}.\footnote{We also used the   analytic continuation to define  $\sum_{m=1}^{\infty}\frac{q^{-m}-q^{m}}{m}=\log(-q)$.}
As a result, 
\be 
\la{x47}
{\rm Z}_1=   c_{0}\,N^{3}\,q^{N-{31\ov 24}} \,  \Big[-q\exp\Big( 7 \sum_{m=1}^{\infty}\frac{1}{m}\frac{q^{m}}{1-q^{m}}\Big)\Big]\  . 
\ee

\subsection{Matching to    large    $N$  correction  in   unrefined    ABJM index} 

Comparing the partition function  \rf{x47} to $q^N \delta\I_{N}^{(1)} $  part of (\ref{2.8})  which in the $\bm{u}=\bm{1}$  case 
takes the form of the  leading  $N^3$  term in \rf{1.2},\rf{2.12}  we observe a close similarity. 
Removing  the  Casimir energy contribution $q^{-{31\ov 24}}$ (as we   are after an
  index that counts, according to the  conjecture of \cite{Arai:2020uwd}, the 
     BPS fluctuations  of  the wrapped M5 brane)\foot{Note that as  there is no Casimir energy term in 3d, 
similar  factor did not appear  in the case of the M2  brane 
supersymmetric partition  function  in \ci{Beccaria:2023sph}.}   
 and assuming that $c_0= {1\ov 6}$   we   match  the  non-trivial $q$-dependent prefactor
 of the "classical" $q^N$  contribution.
  Indeed, we have 
\be
\exp\Big[ \sum_{m=1}^{\infty}\frac{1}{m}\frac{7q^{m}}{1-q^{m}} \Big]
=\exp\Big[  \sum_{m=1}^{\infty}\frac{1}{m}\sum_{m'=1}^{\infty}7q^{m'}\Big] 
= \exp\Big[-7 \sum_{m=1}^{\infty}\log(1-q^{m})\Big]=
\prod_{n=1}^{\infty}(1-q^{n})^{-7}  .\qquad  \la{477}
\ee
This demonstrates that the result of the 
direct computation of the $S^1 \times S^5$  M5 partition function  in AdS$_4 \times S^7$ 
reproduces the $q^N$ prefactor in the ABJM superconformal index in \rf{2.12}. 

A related observation is that the single-particle   partition function 
$\Zsp(q)$ in (\ref{4.1})  matches  the $\bm{u}\to \bm{1}$ limit of  
 the single-particle "M5 brane superconformal index" 
  (\ref{2.11}),   obtained    in \cite{Arai:2020uwd}  by
    an   analytic continuation of the expression  for the  superconformal index of the free (2,0)   6d   multiplet  in flat space
\be\la{488}
\Zsp(q) =   { q^{-1} - 3 q^2 + q^3 + q^4 \ov (1-q)^3} = \Isp^{\rm M5}_{1}(q,\bm{1}) \ .
\ee

\subsection{Generalization to the case  with  non-trivial  $R$-symmetry fugacities $u_{a}$ \la{s4.3}}

In the above discussion we considered  the supersymmetric  partition function 
of   wrapped  $S^1 \times S^5$  M5 brane in  \tAdS$_4 \times S^7$ corresponding to the   contribution to the 
ABJM index \rf{2.3},\rf{2.8},\rf{2.9} 
with $u_1=u_2=u_3=u_4=1$.  
To  include the dependence on  the $SO(8)$   $R$-symmetry  fugacities 
we would need to add extra  $u_a$   dependent twists  in the  $S^7$  angles  in \rf{3.2}. 

Let  us  represent  $S^7$ as a  subspace  $\sum_{a=1}^4  |\rz_a|^2=1$  in $\mathbb C^4$ with  the 
metric  $d{S}_7  =\sum_{a=1}^4  |d\rz_a|^2 =  \sum_{a=1}^4   (dn_{a}^{2}+  n_a^2  d\pps_a^2)$,  $ \sum_{a=1}^4    n_a^2 =1$. 
Adding the   $y$-shifts to  the  angles $\pps_a$   should  reflect the presence of  the $R$-charge   fugacities  $u_a$  in 
the ABJM index  in \rf{2.3} (cf. \rf{3.3}) 
\be \la{144}
d{S}_7  = \sum_{a=1}^4   \big[dn_{a}^{2}+  n_a^2  (d\pps_a  + i \ga_a  dy)^2\big] \ , \ \ \ \ \ \ \ \ \qquad   \sum_{a=1}^4    n_a^2 =1\ . 
  \ee
Here $y \equiv y + \beta $ is the  $S^1_\b \subset  {\rm AdS}_4$ coordinate in \rf{3.2}.
 Taking into account the relative normalizations    of the  factors    in \rf{2.3}  we should  set (note the   constraint  $u_1u_2u_3u_4=1$)
\be \la{145} 
u_a =  e^{-\beta \ga_a}  =   q^{2\ga_a} \ , \ \ \ \ \ \ \qquad   q= e^{-\ha \beta} \ , \qquad \ \ \ \ \ \     \sum^4_{a=1} \ga_a =0 \ . 
\ee
Let us now  assume that M5 brane is wrapping $S^5 \subset S^7$  defined by $\rz_1=0$.
In the parametrization used  in \rf{3.2}  we may identify the angle $z$  with $\pps_1$  and thus 
 have  it shifted as  $dz \to dz + i \ga_1  dy$, with  the wrapped M5    brane solution still described by \rf{39}.
 The  corresponding classical   M5 action   will be given by \rf{3.14}   where the volume   part will not   depend on $\ga_a$ 
 (as   $dy=d \xi^1 $   shifts of   differentials of the $S^5$   will not contribute  the product with $d\xi^1$  corresponding to $S^1_\b$).
 However, now we will get a non-zero contribution from the  WZ term as   $C_6$  in \rf{3.22}  will contain
 $dz \to dz + i \ga_1  dy$   with $v=0$  for  the solution \rf{39} (cf. \rf{314},\rf{3.18})\foot{We assume  a particular orientation on $S^1\times S^5$ so  that $\int  dy \wedge \vol_{S^{5}} = \pi^3 \beta$.}
 \be 
 S_{\rm WZ} =  i T_5 \int C_6 = i {\rm T_5} \int \vol_{S^{5}} \wedge ( i \ga_1  d y)  =    N  \ga_1\beta   \ .   \la{146} 
\ee 
As a result,  we will get  $e^{-S_{\rm WZ} }= e^{-N \ga_1  \beta} =  u_1^N$, 
 reproducing  the overall factor $u_1^N$ in  the  first term in the sum in $\delta\I^{(1)}_{N}(q, \bm{u}) $ in   \rf{2.9}. 

To get  the  factor   $\I^{\rm M5}_{1}(q,\bm{u})$ in \rf{2.9}  we need to  compute  the one-loop correction for 
non-zero  shifts \rf{145}. In view of 
 the  condition $u_1u_2u_3 u_4=1$  in \rf{2.3},
 the simplest   non-trivial  case  to consider is\foot{For  this choice  the  WZ term contribution to the classical action is trivial, $u_1^N=1$.} 
\be \la{444}  u_1=u_4=1, \ \  \qquad \ \ \   u_2= u_3^{-1} =u \equiv q^{2\alpha}  \ , \ \ \ \ {\rm i.e.} \ \ \ \ \ 
\ga_1=\ga_4=0, \ \ \ \ \ga_2=-\ga_3=\alpha \ .  \ee  
The resulting  generalization of the $S^5$ part of the  $S^7$ metric  in \rf{3.2} is then (we rename the coordinates compared to  \rf{144}) 
\be\la{4441}
d{\wtd  S}_5 = dn^{2}_1+dn^{2}_2+dn^{2}_3+n^{2}_1(d\pps_{1}+i\alpha dy)^{2}
+n^2_2(d\pps_{2}-i\alpha dy)^2
+n^2_3 d\pps^2_3  \ , \ \ \ \  \ \ \ \ \ \   n^2_1 + n^2_2 + n^2_3 =1  \ .  \ee
Considering the  analog  of the $S^1_\beta \times S^5$   M5 brane  solution
    in the resulting 
twisted-product space\foot{This space  is  thus  an  analog of the  twisted-product space AdS$_7 \times \wtd S^4$   considered in  connection with the  M2  brane instanton contribution to the  large $N$ 
 superconformal index of  (2,0) theory  in \ci{Beccaria:2023sph}, cf. footnote \ref{f7}.}
 \tAdS$_7 \times \wtd S^7$  we should expect to find that the 
corresponding one-loop partition function  should   have the single-particle  counterpart that 
generalizes \rf{488}  in such a way that it matches  the relevant special  case of \rf{2.11}.
 Namely, we should   find  the following generalization of (\ref{488})
\be 
\la{5.1}
\Zsp(q, u ) =   \frac{q^{-1} -q^{2}(1+u+u^{-1})+q^{3}+q^{4} }{(1-q)(1-u\, q)(1-u^{-1}q)}
=  \Isp_{1}^{\rm M5}(q,1,u,u^{-1},1) \ . 
\ee
To show this one  may   "disentangle" the effect  of the  $J_{12}$-related  (cf. \rf{2.3}) 
fixed shift of the  $S^2\subset$ AdS$_4$ angle  $\varphi$ 
in \rf{3.3}   from  the dependence on the  $\alpha $-parameter  
in \rf{4441}. 
This  is possible  in the static gauge:  when evaluating the  $\nu$-shift   effects on the one-loop   fluctuation  operators   
  due to the $\varphi$ twist in (\ref{3.3}) as in  section 3
 the dependence  on  $\alpha$    can be ignored as   it   can be  absorbed into  the non-fluctuating $S^{5}$ coordinates. 

Then the problem  of finding the dependence on $\alpha$  becomes 
   formally the same as  computing  the  supersymmetric partition function 
of a  single  (2,0)  tensor multiplet   on the  6d space   which is the twisted 
product of $S^1_\beta $ and $ \wtd S^5$  with  the metric \rf{4441}. 
We  describe  this computation in Appendix \ref{app:Rtwist}.
 From  
(\ref{5.400}), (\ref{5.48}), (\ref{5.47})  we get 
the following set  of the 
 $u=q^{2\alpha}$ dependent single-particle partition functions generalizing the 
 $\nu=0$ expressions (\ref{346}), (\ref{354}), (\ref{3.55}) for the scalar, fermion,    and  the antisymmetric tensor  fields 
\ba
\la{m1}
\Zsp_{\phi}(q,0;u) =& \frac{q^{2}(1-q^{2})}{(1-q)^{2}(1-u\,q)^{2}(1-u^{-1}q)^{2}}=\frac{q^{2}(1+q)}{(1-q)(1-u\,q)^{2}(1-u^{-1}q)^{2}}\ ,\\
\Zsp_{\psi}(q,0;u) =& \frac{2c_{\psi}(u)\,q^{5/2}}{(1-q)(1-u\,q)^{2}(1-u^{-1}q)^{2}}\ , \qquad\qquad  c_{\psi}(u) = \tfrac{1}{4}(u+2+u^{-1})\ ,\la{4166} \\
\Zsp_{A}(q;u) =&\frac{q^{3}[c_1(u)-c_2(u)\,q+q^{2}]}{(1-q)(1-u\,q)^{2}(1-u^{-1}q)^{2}}\ ,\la{4177} \\ 
 & c_1(u) = u^{2}+2u+4+2u^{-1}+u^{-2}\ , \qquad \qquad 
   c_2(u) = 2u+1+2u^{-1}\ . \notag
\ea
Using the $u$-dependent partition functions in (\ref{m1}),\rf{4166},\rf{4177}  and 
the  including the   effect  of the $\nu$  shifts  in Table \ref{t1} (which 
amounts to multiplication  of the scalar and fermion contributions by $\frac{1}{2}(q^{2\nu}+q^{-2\nu})$)
 we can check that  total single-particle partition function  indeed reproduces  the expression in  (\ref{5.1}) generalizing \rf{488} to $u\neq 1$
\ba
\Zsp(q,u) =& \Zsp_{\phi}(q,0; u)+2\Zsp_{\phi}(q,1;  u)+2\Zsp_{\phi}(q, \tfrac{3}{2};u)+\Zsp_{A}(q;u)-2\,\Big[\Zsp_{\psi}(q,\tfrac{1}{4};u)
+\Zsp_{\psi}(q, \tfrac{5}{4};u)\Big] \lp
=  \frac{1}{(1-q)(1-uq)(1-u^{-1}q)}\Big[q^{-1}-q^{2}(1+u+u^{-1})+q^{3}+q^{4}\Big] \ . \la{4888}
\ea
Note that expanding \rf{4888}  in $q$  we get 
\ba
\la{5.53}
\Zsp(q,u) =q^{-1}+u+1+u^{-1}+\mc O(q).
\ea
The $q$-independent   part  $u+1+u^{-1} $ here  
 is  the $u\neq 1$    generalization of  the term  3  in (\ref{4.1})   which, as we discussed  above,  leads to a  formal divergence in $\bar F$ in \rf{x41} that   should  be attributed to the contribution of the zero modes. 
 Since $\sum_{m=1}^\infty  {1\ov m} (u^m   + u^{-m}) =-  \log [(1-u)(1-u^{-1})]$, here  the  contributions of $u+ u^{-1}$  can be formally regularized using the analytic continuation in $u$.  The divergence due to the  remaining term 1  in \rf{5.53}  should be associated to the contribution of two bosonic zero modes  that  should  produce  a prefactor  $(\sqrt{\rm T_5})^2\sim N$  (cf. \rf{x477}). 
 
 Indeed, analyzing  the expression for the  ABJM  index (\ref{2.9}) in the limit of $\bm{u}\to (1,u,u^{-1},1)$   and large $N$ 
  (generalizing the discussion in Appendix \ref{app:finiteN} in  the $\bm{u}\to {1}$ case  leading to \rf{1.2},\rf{2.12})  we get\footnote{There are  also 
   terms proportional to $u^{N}$ and $u^{-N}$ but they  have subleading coefficients $\sim N^{0}$.}
\be\la{4299}
\I_{N}^{\rm ABJM}(q,u) = \I_{\rm KK}^{\rm ABJM}(q,u) \, \Big[1-\frac{N}{(1-u)(1-u^{-1})}\,\GG(q,u)\,q^{N}+ \mc O(N^{0}q^{N})\Big] \ . 
\ee
Thus  the leading large $N$  contribution 
 has the expected factor $\frac{N}{(1-u)(1-u^{-1})}$    associated with the presence of the $q$-independent 
 term in \rf{5.53}. 
The  function $\GG(q,u)$  in \rf{4299}   is given by  
(here $n_{123}\equiv n_{1}+n_{2}+n_{3}$)
\ba
\GG(q,u) =& -\frac{q}{1-q}\mathop{\prod_{n_{1},n_{2},n_{3}\ge 0,\,  n_{123}\geq 2}}
^{\infty}\frac{1}{1-q^{n_{123}-1}u^{n_{2}-n_{3}}} \lp\ \ \ 
\times\prod_{n_{1},n_{2},n_{3}=0}^{\infty}\frac{
(1-q^{n_{123}+2}u^{n_{2}-n_{3}})
(1-q^{n_{123}+2}u^{n_{2}-n_{3}+1})
(1-q^{n_{123}+2}u^{n_{2}-n_{3}-1})
}{
(1-q^{n_{123}+3}u^{n_{2}-n_{3}})
(1-q^{n_{123}+4}u^{n_{2}-n_{3}})
}.\la{4211}
\ea
Switching on  all 4   non-trivial   fugacities  $u_a$ we should find that the  zero mode  contribution is 
completely regularized\foot{Note that the small $q$ expansion of \rf{2.11} is 
$\Isp^{\rm M5}_{1}(q,\bm{u}) = q^{-1}   u_1^{-1} + ( u_2 + u_3 + u_4) u_1^{-1} + {\cal O} ( q) $ (cf. \rf{5.53}).}
 and,  correspondingly, 
  there is no $N$-dependent prefactor in the analog of the $q^N$ term  in \rf{4299}.

\section{Supersymmetric partition function of  free  $(2,0)$  multiplet on $S^{1}\times S^{5}$ }

The   expression for the superconformal index  
of a free 
$(2,0)$ tensor multiplet    in  flat 6d space  \cite{Kim:2012ava} 
   was  used in  \cite{Arai:2020uwd}     to find the  "M5 brane index"  \rf{2.11}   by  an analytic continuation 
and re-identification of fugacities.  
For completeness, let us    show how  obtain the index of the   $N=1$ (2,0)  theory    by directly computing 
the  supersymmetric partition function  of this   6d tensor multiplet 
 on $S^1 \times S^5$.

The simplest (unrefined)  version of the (2,0)     superconformal index   is defined  as 
$ \Tr\big[(-1)^{\rm F}\,  q^{H-R_{12}}\big]$   where   $R_{12}$   is a    generator  of the 
 $R$-symmetry $SO(5)$ group. Hence to preserve  supersymmetry one is to  include a  particular  flat connection  in the scalar and fermion 
 kinetic operators on $S^1 \times S^5$  corresponding to a rotation in  the "target-space"   12-plane    \cite{Kim:2012ava}.
 Like in the  discussion  in section 3  this   will lead to particular shifts  of the $S^1$ mode numbers   in the "mass" terms in 
 the  corresponding   differential operators on $S^5$  (cf. \rf{3.40}).
 
As a result,   we   get   three    scalars with no shift,   two    scalars   shift $\nu=\frac{1}{2}$ and  
 four   MW  6d fermions  with  shift $\nu=\frac{1}{4}$.\foot{Here the   length  of $S^1$ is 
$\beta$, not $\ha \beta$, and the shift is   still relative to $\nb= {2\pi \ov \b}$.}  
 Using  the data in  Table  \ref{t1}  we then  find (cf. \rf{4.1},\rf{4.2}) 
\ba
\Zsp(q) =& 3\Zsp_{\phi}(q,0)+2\Zsp_{\phi}(q,\tfrac{1}{2})+\Zsp_{A}(q)
-4\, \Zsp_{\psi}(q,\tfrac{1}{4}) = \frac{q}{1-q}, \la{414}\\
E_{c,\rm tot} =& 3E_{c,\phi}(0)+2E_{c,\phi}(\tfrac{1}{2})+E_{c, A}-4 E_{c,\psi}(\tfrac{1}{4}) = -\tfrac{1}{24}\ .\la{415}
\ea
These expressions  agree with  the ones for the 
 Schur index  (and   supersymmetric Casimir energy)  
  for a single 6d $(2,0)$ tensor multiplet    (see also  (\ref{C.12})) 
  obtained in \cite{Kim:2012ava}  by counting  BPS states with $R$-charge dependent weights.
  

We  may  also  consider a generalization  to the case of  two $SO(5)$ 
 $R$-symmetry   generators 
 \be
Z_{\eta}(q) = \Tr\Big[(-1)^{\rm F}\,  q^{H-R_{\eta}}\Big],\qquad \qquad 
R_{\eta}=\frac{\eta}{2}R_{12}+\frac{1-\eta}{2}R_{34} \ ,   \la{4145}
\ee
where $\eta$  is a   free parameter  (denoted by $\Delta$ in \cite{Kim:2012ava}).  
In this case  there are two orthogonal components in the  flat connection (and two  $\Gamma$-matrix   projectors
in the  spinor covariant derivative).
As a result we get 
 one  scalar with $\nu=0$, two with $\nu=\frac{1}{2}\eta$  and   two  with $\nu=\frac{1}{2}-{\eta\ov 2}$  as well as 
two   fermions   with   $\nu=\frac{1}{4}$ and two   with  $\nu=\frac{1}{4}-\frac{\eta}{2}$. 
The resulting single-particle partition function is   
\ba
\la{4.12}
\Zsp_{\eta}(q) =& \Zsp_{\phi}(q,0)+2\Zsp_{\phi}(q,\tfrac{\eta}{2})+2\Zsp_{\phi}(q,\tfrac{1}{2} -\tfrac{\eta}{2})
+\Zsp_{A}(q)-2\,\Big[\Zsp_{\psi}(q,\tfrac{1}{4})+\Zsp_{\psi}(q,\tfrac{1}{4}-\tfrac{\eta}{2})\Big] \lp
= \frac{q^{1+\eta} + q^{2-\eta} -3q^{2}+q^{3}}{(1-q)^{3}}\ .
\ea
Note that  for $\eta=-2$  and $\eta=3$  this coincides   with   \rf{4.1}\rf{488} (see also below). 

Comparing \rf{4.12}  to   the general   expression for the  single-particle  superconformal 
 index  of  one     $(2,0)$  multiplet 
$\Isp_{N=1}^{(2,0)}(q,\bm{y}, u)$ in \rf{222},(\ref{C.3}),(\ref{C.4}) we  conclude that 
\be
\la{4.13}
\Zsp_{\eta}(q) = \Isp_{N=1}^{(2,0)}(q^{\frac{3}{4}}, 1, 1, 1, q^{\frac{1}{2}-\eta}) \ , \ee 
 which corresponds
  to   ${\rm I}_{N=1}^{(2,0)}(q^{\frac{3}{4}}, 1, 1, 1, q^{\frac{1}{2}-\eta})  = \Tr\Big[(-1)^{\rm F}\,  q^{\frac{3}{4} [H+\frac{1}{3}(J_{12}+J_{34}+J_{56})]
 +(\frac{1}{2}-\eta)(R_{12}-R_{34})}\Big].$
Using that  $\Delta$ in (\ref{C.2})  should be set to 0 in \rf{C.3},  this   simplifies  indeed to \rf{4145}. 
Combining (\ref{4.13}) with the analytic continuation rule (\ref{2.10}) 
that  gives $\Isp_{1}^{\rm M5}(q, \bm{u})$  in (\ref{2.11}) the relation  in 
\rf{4.13} may be written also as  
\be
\Zsp_{\eta}(q) = \Isp_{1}^{\rm M5}(q^{\frac{1+\eta}{4}}, q^{\frac{3\eta -9}{4}}, q^{\frac{3-\eta}{4}}, q^{\frac{3-\eta}{4}}, q^{\frac{3-\eta}{4}})\ . \la{4147}
\ee
Thus  for  $\eta=3$  it should reproduce the "unrefined"   $\bm{u}\to \bm{1}$ limit  of  the M5 brane index. 
 Indeed, for $\eta=3$   the  values of the  individual shifts   and  the total expression in (\ref{4.12}) 
   become the same as \rf{4.1} 
 that  we found  above   by 
 the  direct  analysis of fluctuations of  the wrapped M5 brane  in \tAdS$_4 \times S^7$.
  

\iffa
\section{Conclusions}
We remark that the counting approach does not  immediately apply to scalars and fermions as there we  have effects of target space twisting 
and 3-form coupling (via $C_6$ in scalars  case  and $F_4$ coupling in the fermion case). So, in those cases, 
we needed to do a direct computation  starting with determinants of fluctuation operators. 
\fi

\section*{Acknowledgements}
We are grateful  to S. Giombi for an initial    collaboration  
and   many useful discussions. 
MB was supported by the INFN grants GSS and GAST.
AAT was supported in part by the STFC Consolidated Grants ST/T000791/1 and ST/X000575/1.


\appendix

\section{Brane instanton expansion of superconformal index}
\la{app:brane-exp}

In a superconformal theory with an AdS dual  the large  gauge group rank $N$   corrections to superconformal index 
may be interpreted  in terms of contributions of  brane instantons.
Considering for simplicity the dependence on a single fugacity $q$, the structure of  $N$ dependence is encoded in the 
expansions  like   (\ref{1.1}), (\ref{1.2}), i.e. \footnote{More generally one has $q^{kN}\to q^{a_{k}N}$ where $a_{k}$ is some increasing  sequence of positive integers.}
\be
\la{B.1}
\I_{N}(q) = I_{\infty}(q)\, \Big[1+\sum_{k}q^{kN}\delta\I_{k}(q)\Big],
\ee
where $\delta\I_{k}(q)$ is regular at  small $q$ and the only explicit dependence on  $N$ is in the factor $q^{kN}$. 
Such  structure is observed in many  special cases \cite{Arai:2019wgv,Arai:2019aou,Arai:2020qaj,Imamura:2021dya,Fujiwara:2023azx}. For historical reasons 
it is usually  called "giant graviton expansion" (see, e.g., \cite{Gaiotto:2021xce})
 although   a more appropriate name would  be a   "brane instanton expansion".

The paradigmatic case  is  that of  the  4d  
$\N=4$ $U(N)$ SYM  theory   corresponding to a system of  $N$  coincident  D3 branes in type IIB superstring model.
The large  $N$  limit of its index  $\I_{\infty}(q)$ counts  the  $S^5$ 
Kaluza-Klein   BPS   states of AdS$_{5}\times S^{5}$ supergravity.  
The structure of finite $N$ corrections    from the state counting  perspective  is   non-trivial 
 \cite{Biswas:2006tj,Mandal:2006tk,Kim:2006he,Lee:2023iil}. They are 
important, e.g.,  for the corresponding BPS black hole   entropy   count
 since the asymptotic growth of the index  (\ie the number of states at increasing charge)
is faster than that of a  gas of the KK modes \cite{Cabo-Bizet:2018ehj,Choi:2018hmj,Benini:2018ywd,Murthy:2022tbj,Agarwal:2020zwm}.
Finite $N$ corrections to the index  take  the form (\ref{B.1}) where each term in the square
 brackets is the effect of $k$  wrapped D3 branes   (or  "giant gravitons" 
\cite{McGreevy:2000cw,Grisaru:2000zn,Hashimoto:2000zp,Mikhailov:2000ya,Balasubramanian:2001nh}).

The functions $\delta\I_{k}(q)$ can be computed 
by considering  branes multiply wrapped  around topologically trivial 3-cycles in the internal 
$S^{5}$ space \cite{Imamura:2021ytr}. 
If we represent  $S^{5}$ as  $|\rz_{1}|^{2}+|\rz_{2}|^{2}+|\rz_{3}|^{2}=1$,  the  three 3-cycles  are defined by $\rz_{i}=0$. Denoting the wrapping numbers by 
$(n_{1},n_{2},n_{3})$,  the  theory on the wrapped D3 branes  is  $U(n_{1})\times U(n_{2})\times U(n_{3})$ gauge theory
with bi-fundamental multiplets in a ring quiver diagram. 
Here  $q^{kN}$ in \rf{B.1} 
is   given by $q^{n_{1}N}q^{n_{2}N}q^{n_{3}N}$ 
and represents the total classical prefactor coming from the 
classical charges and energy of the wrapped brane system. Also, we get\footnote{In general, at mathematical level, 
the expansions like (\ref{B.1})  are not unique. For instance, an exact 
expansion for the $\mc N=4$ SYM index was given in \cite{Murthy:2022ien}, but it differs from the ones arising from wrapped D3 branes \cite{Liu:2022olj,Eniceicu:2023uvd}.} 
\be\la{b2}
\delta\I_{k}(q) = \mathop{\sum_{n_{1},n_{2},n_{3}=0}}_{n_{1}+n_{2}+n_{3}=k}^{\infty}\delta\I_{n_{1},n_{2},n_{3}}(q) \ . 
\ee
The actual 
evaluation of the so-called "brane indices"  $\delta\I_{n_{1},n_{2},n_{3}}(q)$ goes in two steps:
(a) first, one finds  the single-letter index $\delta\Isp_{n_{1},n_{2},n_{3}}(q)$ of the brane world-volume superconformal theory;\foot{In favourable cases, 
this may be  accomplished by  an 
analytic continuation of the index of the  superconformal theory in flat space with the same superconformal algebra.}
(b) 
second, to get the brane index, one has to integrate the plethystic exponential $\PE [\delta\Isp_{n_{1},n_{2},n_{3}}]$ over the $U(n_{1})\times U(n_{2})\times U(n_{3})$ gauge  holonomies.

The second step 
 is far from trivial because it is unclear which contour should be used to integrate the holonomies for the 
analytically continued single-particle index.
 For instance,  representing the holonomies as $U(1)$ phases $w_{i}$, it is known 
that  the standard cycles $|w_{i}|=1$ give a wrong result.  
In some cases, it is possible to match the finite $N$ index by some  special {\em ad hoc} 
choices, at least up to some total wrapping level $k$. 
A prescription working up to $k=3$ was proposed in  \cite{Imamura:2021ytr}
and other examples treated by the same approach can be found in 
\cite{Arai:2019wgv,Arai:2019aou,Arai:2020uwd,Arai:2020qaj,Imamura:2021dya,Fujiwara:2023azx}.
For recent discussions of the analytical continuation for generic wrapping see \cite{Gaiotto:2021xce,Imamura:2022aua,Lee:2022vig,Beccaria:2023zjw,Fujiwara:2023bdc}.

\section{Superconformal index of  $(2,0)$  theory 
and its Schur limit \la{aC}}

The 6d $(2,0)$ superconformal algebra is $\mk{osp}(8^{*}|4)$ with the bosonic subalgebra  $\mk{so}(2,6)\oplus \mk{so}(5)$
having  six Cartan generators
\be
\bm{\mc C} = (H, \ J_{12}, \  J_{34}, \  J_{56}, \  R_{12}, \  R_{34})\ .
\ee
The superconformal index  of the $(2,0)$ theory  is defined in terms of a supercharge $Q$  satisfying 
$[\mc C_{I},Q] = \frac{1}{2}\sigma_{I}\,Q$, $I=1, \dots, 6$ with ${\sigma}_I = (1,-1,-1,-1,1,1)$.
The subalgebra commuting with $Q$ is $\mk{osp}(6|2)\oplus \mk{u}(1)_{\Delta}$ with the  bosonic
algebra $\mk{su}(1,3)\oplus \mk{su}(2)$ and the central factor $\mk{u}(1)_{\Delta}$ having the  generator
\be
\la{C.2}
\Delta = \{Q,Q^{\dagger}\} = H-(J_{12}+J_{34}+J_{56})-2(R_{12}+R_{34})\ .
\ee
The  superconformal index  is then 
\be
\la{C.3}
\I^{(2,0)}_{N}(q, \bm{y}, u) = \mathop{\Tr}_{\Delta=0}\Big[(-1)^{\rm F}\,  q^{H+\frac{1}{3}(J_{12}+J_{34}+J_{56})}\,
 y_{1}^{J_{12}}\, y_{2}^{J_{34}}\, y_{3}^{J_{56}}\, u^{R_{12}-R_{34}}\Big],\qquad y_{1}y_{2}y_{3}=1 .
\ee
The trace is restricted to the states with $\Delta=0$ (contributions of states with $\Delta>0$ cancel  in pairs).



The  Schur  limit of the   index \rf{C.3} is defined by  imposing the condition on  the fugacities
\be
\la{C.5}
q^{\frac{2}{3}}\,u\,y_{1}^{-1}=1 \ , 
\ee
that   implies invariance under an    additional supercharge $Q'$
satisfying $[\mc C_{I},Q'] = \frac{1}{2}\sigma_{I}\,Q'$, $I=1, \dots, 6$ with ${\sigma}_I = (1,-1,1,1,1,-1)$. 
 It is associated with the 
second $\mathfrak{u}(1)$ with  the generator
\be
\la{C.6}
\Delta' = \{Q',Q'^{\dagger}\} = H-(J_{12}-J_{34}-J_{56})-2(R_{12}-R_{34}).
\ee
A convenient parametrization for the independent fugacities (obeying (\ref{C.5}) and $y_{1}y_{2}y_{3}=1$) is 
\be
\la{C.7}
q = q'\, x', \quad y_{1} = q'^{\frac{2}{3}}\,x'^{-\frac{4}{3}}, \qquad 
y_{2} = q'^{-\frac{1}{3}}\,x'^{\frac{2}{3}}\,y, \qquad y_{3}=q'^{-\frac{1}{3}}\,x'^{\frac{2}{3}}\,y^{-1}  \ , 
\qquad u = x'^{-2} \ . 
\ee
In this limit  the index \rf{C.3} reads
\be\la{b8}
\I_{N}^{(2,0)}(q', y) = \mathop{\Tr}_{\Delta=\Delta'=0}\Big[(-1)^{\rm F}\, q'^{H+J_{12}}\,y^{J_{34}-J_{56}}\Big].
\ee
From the condition $\Delta = \Delta' = 0$  we get $H+J_{12} = 2\,(H-R_{12})$. 
As a result, the unrefined ($y=1$)  Schur index   is given by 
\be \la{C.10}
\I^{(2,0)}_{N}(q')\equiv \I^{(2,0)}_{N}(q', 1) = \mathop{\Tr}_{\Delta=\Delta'=0}\Big[(-1)^{\rm F}\,  q'^{2(H-R_{12})}\Big] \ , \qquad  \ \ \ q'^{2} = e^{-\beta}  \ . 
\ee
The  leading   large  $N$ correction  \rf{B.1} to this index    scales as  $\sim q'^{2N} = \exp(-\beta N)$
 \cite{Arai:2020uwd}. 
 
  The Schur index $\Tr[(-1)^{\rm F}e^{-\beta(H-R_{12})}]$   may be computed as  a   supersymmetric 
 partition function on $S^{1}_{\beta}\times \wtd S^{5}$ with periodic fermions and $R$-charge related twist
   (cf. \cite{Aharony:2021zkr}).
 As was demonstrated    in \cite{Beccaria:2023sph},  
 the  leading non-perturbative contribution to the large $N$  expansion   of 
  this index 
may   represented in the dual  M-theory  description 
as  the  semiclassical    partition function  of M2  brane  in 
AdS$_{7}\times \wtd S^{4}$ where $\wtd S^{4}$
 has  one angle  mixed   ($z\to z+iy$)  with   the  coordinate of $S^1_\beta \subset$ 
AdS$_{7}$.

Let us   note  also that 
for $N=1$    (2,0) tensor multiplet  the single-particle index corresponding to \rf{C.3} 
  reads  (see, e.g.,  \cite{Bhattacharya:2008zy,Arai:2020uwd})\footnote{This expression agrees with Eq.(3.35) in \cite{Bhattacharya:2008zy}, with   a  suitable identification of  the fugacities, i.e.  $x=q^{1/3}, u=z^{1/2}$, etc.}
\be
\la{C.4}
\Isp^{(2,0)}_{N=1}(q,\bm{y},u) = \frac{q^{2}(u+u^{-1})-q^{\frac{8}{3}}(y_{1}^{-1}+y_{2}^{-1}+y_{3}^{-1})+q^{4}}{(1-q^{\frac{4}{3}}y_{1})(1-q^{\frac{4}{3}}y_{2})(1-q^{\frac{4}{3}}y_{3})}.
\ee
In  the "unrefined"   case   $\bm{y},u\to 1$   it  is given by 
\be\la{b10}
\Isp_{N=1}^{(2,0)}(q) =\Isp_{N=1}^{(2,0)}(q,\bm{1},1) =\frac{2q^{2}-3q^{\frac{8}{3}}+q^{4}}{(1-q^{\frac{4}{3}})^{3}}.
\ee
To find its   Schur limit, we change  the variables as in (\ref{C.7})  and set  $x'=y=1$ getting
\be
\la{C.12}
\Isp_{N=1}^{(2,0)\, \rm Schur}(q') =\frac{q'^{2}}{1-q'^{2}}.
\ee
 Let us    add a comment on a comparison  of the  prefactors  of the $q^N$ terms in the  expansion  of the indices 
 of the (2,0)     \rf{1.1}      and the ABJM \rf{1.2} theories.
As was reviewed in the Introduction,
 the leading large $N$ correction to the superconformal index of the (2,0) theory  in \rf{1.1}  
 has a  prefactor   $-{ q\ov (1-q)^2}$ 
  with a  simple dependence 
 on $q$. 
 At the same time, the counterpart of this  prefactor  in the ABJM  index  is  proportional to  the  $G_0(q)$ 
 function in  \rf{1.2} that 
 has a  complicated  dependence on $q$. 
 This may be puzzling as both  were shown to  originate from the one-loop partition functions -- 
 of the  $S^1 \times S^2$ 
 M2  brane in the  twisted  AdS$_7 \times S^4$ background  in the case of \rf{1.1} 
 \cite{Beccaria:2023sph} 
 and of  the $S^1 \times S^5$  M5 brane  in the twisted AdS$_4 \times S^7$ in the case of \rf{1.2} (see section 4). 
 
 One may  understand  the reason for the simplicity of the prefactor in \rf{1.1} 
 from  its alternative   derivation   in \cite{Arai:2020uwd}. 
According to the proposal in \cite{Arai:2020uwd},   an indirect   way to  compute   this  prefactor 
is to     exploit   the isomorphism between  the   superconformal  algebras of the two 3d models:  
 the  quadratic fluctuation  action  for   the wrapped M2 brane  in curved AdS$_7 \times S^4$ target space and  of 
 a   single  $\N=8$   3d scalar multiplet in flat space  (or, equivalently, of the  $N=1$ ABJM  model)  
 defined on $S^{1}\times S^{2}$. 
 For the latter theory, the expression for  the  index with  non-trivial 
 chemical potentials reads\footnote{See Eqs.(40) and (17) in \cite{Arai:2020uwd}. 
 This  expression  reduces to the $N=1$ case in (\ref{2.6}) for $\bm{u}=1$.}
\be
\I_{N=1}^{\rm ABJM}(q, \bm{u}) = \PE\Big[\frac{q(u_{1}+u_{2}+u_{3}+u_{4})-q^{3}(u_{1}^{-1}+u_{2}^{-1}+u_{3}^{-1}+u_{4}^{-1})}{1-q^{4}}\Big].
\ee
To relate   fugacities of  this  model 
 to those  of the (2,0) theory index we may  still   use 
(\ref{2.10}).\foot{The   isomorphism  used  in \cite{Arai:2020uwd}   is between  the 
(i)  unbroken part of the superconformal algebra on the wrapped branes (M2 or M5 in the two cases), and 
(ii) the subalgebra respecting the supercharge used to define the corresponding  index
  (\cf also footnote \ref{fff1}).}
Explicitly,  in the case of the  index of the  ABJM  model on $S^1 \times 
 S^3$      {\it and}  the index of  M5  brane  wrapped on $S^1 \times  S^5$  this is  $su(2|4) \times  u(1)$. 
In the case of the index of the   (2,0)  theory on $S^1 \times  S^5$  {\it and} 
  M2   brane wrapped on $S^1 \times  S^3$   this is $su(4|2) \times  u(1)$. 
The  observation of  \cite{Arai:2020uwd}  is that  $su(2|4) \times  u(1)$   and $su(4|2) \times  u(1) $ are isomorphic.}
     Adopting the Schur parametrization (\ref{C.7}),  we then get 
\be
(q,u_{1},u_{2},u_{3},u_{4}) = (q'^{\frac{1}{2}}x', q'^{-\frac{5}{2}}x'^{-1}, q'^{\frac{3}{2}}x'^{-1}, q'^{\frac{1}{2}}x' y, q'^{\frac{1}{2}}x'y^{-1}).
\ee
In the unrefined Schur limit $x' = y=1$ this reads 
\be
\la{4.7}
(q,u_{1},u_{2},u_{3},u_{4}) = (q'^{\frac{1}{2}}, q'^{-\frac{5}{2}}, q'^{\frac{3}{2}}, q'^{\frac{1}{2}}, q'^{\frac{1}{2}}).
\ee
Using (\ref{4.7})  and assuming as  usual an   analytic continuation in computing  the plethystic exponential,   we get 
\be
\la{x412}
\I_{N=1}^{\rm ABJM}(q') = \PE[q'^{2}+q'^{-2}] = -\frac{q'^{2}}{(1-q'^{2})^{2}} \ . 
\ee
This is  the same as  the prefactor in  (\ref{1.1}), taking into account  the definition of the (2,0) index in (\ref{C.10}).


\section{Leading large $N$  correction  
to  ABJM index in  unrefined limit}
\la{app:finiteN}

Here we will   discuss the   leading term  $\delta \I^{(1)}_{N}(q,\bm u)$   in \rf{2.8}  and    large $N$   expansion 
in  the unrefined limit $\bm u\to 1$   deriving \rf{2.12},\rf{2.13}. 

To  find  $\delta \I^{(1)}_{N}(q,\bm{u})$  we need to compute  $\PE[\Isp^{\rm M5}_{a}(q,\bm{u})]$ in  (\ref{2.9}) (see, e.g., 
 \cite{Imamura:2021ytr}). Let us focus on  the   coefficient $\PE[\Isp^{\rm M5}_{1}]$ of the  $u_1^N$  term  in  (\ref{2.9}).
 Expanding  it  at small $q$ gives
\ba
\Isp^{\rm M5}_{1}(q, \bm{u}) =& \frac{1}{u_1 
q}+\frac{u_2}{u_1}+\frac{u_3}{u_1}+\frac{u_4}{u_1}+\Big(\frac{u_2^2}{u_1}+\frac{u_2 u_3}{u_1}+\frac{u_3^2}{u_1}+\frac{u_2 
u_4}{u_1}+\frac{u_3 u_4}{u_1}+\frac{u_4^2}{u_1}\Big) q \lp
+\Big(-\frac{1}{u_1 
u_2}+\frac{u_2^3}{u_1}-\frac{1}{u_1 u_3}+\frac{u_2^2 
u_3}{u_1}+\frac{u_2 u_3^2}{u_1}+\frac{u_3^3}{u_1}-\frac{1}{u_1 
u_4}+\frac{u_2^2 u_4}{u_1}+\frac{u_2 u_3 u_4}{u_1}\lp
+\frac{u_3^2 
u_4}{u_1}+\frac{u_2 u_4^2}{u_1}+\frac{u_3 
u_4^2}{u_1}+\frac{u_4^3}{u_1}\Big)\, q^2+\mc O(q^3). 
\ea
The  terms with  positive/negative coefficients correspond to the contribution of bosonic/fermionic  BPS states. 
The full plethystic exponential \rf{2.4}   may be written as  a product of contributions of   monomials 
\be
\PE [\Isp^{\rm M5}_{1}(q, \bm{u})] = \PE\Big[\frac{1}{u_1 q}\Big]\ \PE\Big[\frac{u_2}{u_1}\Big]\ \cdots.
\ee
For a single monomial with coefficient $\pm 1$, one  has 
\be
\PE[\pm {u}^{{a}}q^{b}] = (1-{u}^{{a}}q^{b})^{\mp 1},
\ee
which is to be understood in terms of  an analytic continuation in the fugacities.\footnote{If the coefficient of a monomial is not $\pm 1$, we use that  $\PE[2X] = (\PE[X])^{2}$, etc.}
As a result, we get 
\ba
\PE[\Isp^{\rm M5}_{1}(q, \bm{u})] =  &  -\frac{u_1^4}{(u_1-u_2) (u_1-u_3) (u_1-u_4)}\,q \no \\   &-\frac{u_1^3 
\big[u_1^2+u_2^2+u_3^2+u_3 u_4+u_4^2+u_2 (u_3+u_4)\big] }{(u_1-u_2) 
(u_1-u_3) (u_1-u_4)}\,q^{2}+\mc O(q^{3}).
\ea
The unrefined limit $\bm{u}\to 1$  appears to be  singular, but, in fact,  the poles cancel after
 summing over $a=1, \dots, 4$ in (\ref{2.9})
and one   may check  the  agreement with the explicit expansions in (\ref{2.6}) for low  values  of $N$. 

It is important to note  that 
 the expansion coefficients in  the unrefined  expression  for 
 $\delta\I^{(1)}_{N}(q)$  in \rf{2.6} depend on $N$. This is due to the 
factors $u_{a}^{N}$ that produce an  $N$-dependent leftover after the  pole cancellation.\foot{Such a degeneracy enhancement 
 at special points in the fugacity space is a  common  phenomenon  (see,   for instance,  Eq. (4.4) in \cite{Lee:2022vig}  and also \cite{Beccaria:2023zjw}).}  Indeed,   we find that for generic $N$
\ba
\delta\I^{(1)}_{N}(q) =& -\tfrac{1}{6}(N+2)(N+3)(N+4)\,q-\tfrac{1}{6}(N+1)(N+3)(7N+8)\,q^{2}\lp
-\tfrac{5}{6}(N+1)(N+2)(7N-12)\,q^{3}-\tfrac{10}{3}(N+1)(7N^{2}-25N-9)\,q^{4}+\cdots.
\ea
Thus  for large $N$  the  leading  enhancement in degeneracy  is $N^{3}$  one  
\ba
\delta\I^{(1)}_{N}(q) =& \tfrac{1}{6}\,N^{3}\,G_{0}(q)+\mc O(N^{2}), \\
G_{0}(q) =& -q-7q^{2}-35q^{3}-140q^{4}-490q^{5}-1547q^{6}-4522q^{7}-12405q^{8}-32305q^{9}+\cdots
\ea
To determine the closed form of $G_{0}(q)$,  we   go back to  the  general 
expression for $\Isp^{\rm M5}_{1}(q,\bm{u}) $ in \rf{2.11} 
\ba
\Isp^{\rm M5}_{1}(q,\bm{u}) =& \sum_{n_{2},n_{3},n_{4}=0}^{\infty}q^{n_{2}+n_{3}+n_{4}}u_{2}^{n_{2}} u_{3}^{n_{3}} u_{4}^{n_{4}}\Big[
q^{-1}u_{1}^{-1}-q^{2}u_{1}^{-1}(u_{2}+u_{3}+u_{4})+q^{3}u_{1}^{-1}+q^{4}\Big] .
\ea
Then  (here $n_{234}\equiv n_{2}+n_{3}+n_{4}$)
\ba
& \PE [\Isp^{\rm M5}_{1}] \lp
= \prod_{n_{2},n_{3},n_{4}=0}^{\infty}\frac{
(1-q^{n_{234}+2}u_{1}^{-1}u_{2}^{n_{2}+1}u_{3}^{n_{3}}u_{4}^{n_{4}})
(1-q^{n_{234}+2}u_{1}^{-1}u_{2}^{n_{2}}u_{3}^{n_{3}+1}u_{4}^{n_{4}})
(1-q^{n_{234}+2}u_{1}^{-1}u_{2}^{n_{2}}u_{3}^{n_{3}}u_{4}^{n_{4}+1})
}{(1-q^{n_{234}-1}u_{1}^{-1}u_{2}^{n_{2}}u_{3}^{n_{3}}u_{4}^{n_{4}})
(1-q^{n_{234}+3}u_{1}^{-1}u_{2}^{n_{2}}u_{3}^{n_{3}}u_{4}^{n_{4}})
(1-q^{n_{234}+4}u_{2}^{n_{2}}u_{3}^{n_{3}}u_{4}^{n_{4}})
}\, .
\ea
The pole at $\bm{u}\to \bm{1}$ comes from the first factor in the denominator when $n_{234}=1$. Examining the residue we obtain 
\ba
G_{0}(q) =&\mathop{\prod_{n,m,\ell=0}^{\infty}}_{n+m+\ell\neq 1}\frac{1}{1-q^{n+m+\ell-1}}\, \prod_{n,m,\ell=0}^{\infty}\frac{(1-q^{n+m+\ell+2})^{3}}{(1-q^{n+m+\ell+3})(1-q^{n+m+\ell+4})} \no \\
=&\frac{1}{1-q^{-1}}\prod_{n=2}^{\infty}\Big[\frac{1}{1-q^{n-1}}\Big]^{\frac{(n+1)(n+2)}{2}}\, \prod_{m=0}^{\infty}\Big[\frac{(1-q^{m+2})^{3}}{(1-q^{m+3})(1-q^{m+4})}\Big]^{\frac{(m+1)(m+2)}{2}} \ . 
\ea
After some simplification this gives  the expression in \rf{2.13}
\be
\la{D.15}
G_{0}(q) = -q\,	\prod_{n=1}^{\infty}\frac{1}{(1-q^{n})^{7}} = -q^{\frac{31}{24}}\,\big[\eta(q)\big]^{-7} \   , \qquad \qquad 
\eta(q) = q^{\frac{1}{24}}\prod_{n=1}^{\infty}(1-q^{n})    \ . 
\ee
Note that  the $q^{\frac{31}{24}}$ factor   here  happens to be  the inverse of the supersymmetric 
Casimir  energy factor in the one-loop M5 brane partition function in   \rf{x477}. 


\section{Free energy of  a conformal scalar field on $S^{1}\times S^{5}$}  
\la{app:free-energy}

Here we present  the  computation  of the free energy (\ref{3.48}) of the conformally coupled  massless 6d  scalar 
on $S^{1}_\beta \times S^{5}$  with an extra  shift $\nu$  of  the $S^1$ mode number $\nb = {2\pi n\ov \beta}$. That shift 
 may be    due   to the presence of a flat connection  or $y=\xi^1$ dependent rotation of a complex scalar. 
Since $S^{1}\times S^{5}$ is  conformally flat,   there are no logarithmic UV divergences in the  corresponding  
one-loop  effective action or free energy.  One  should  still  subtract  power divergences  using
  $\zeta$-function  regularization.

For the   case of a conformally coupled  massless scalar on $S^1 \times S^5$  
  one has (see  also  footnote \ref{foot:lapl})\foot{To recall, 
compared to the  standard normalization 
we are considering a scalar field on $S^1 \times S^5$ 
with an extra $1\ov 4$ in the metric \rf{40}, i.e. with the
 effective length of  the "temperature circle"  $S^1$   being $\bar \beta=\ha \beta$.}  
\ba\la{dd1}
&F = \tfrac{1}{2} \sum_{n=-\infty}^\infty \sum_{k=0}^{\infty} \dd_k \, \log  \lambda_{n,k} \ , \ \ \ \ \ \qquad \qquad 
 \dd_k= \tfrac{1}{12} (k+1)(k+2)^{2}(k+3) \ , \ \ \ \ \\
&  \l_{n,k}=  ({2 \pi n\ov \bb} )^2  +   \omega^2_k \ , \ \ \ \ \ \ \ \  \omega_k =k+2  \ , \qquad\qquad  \bb\equiv  \tfrac{1}{2} \beta \ . 
  \la{d0}
\ea
One may define the  spectral   zeta function
   $
    \zeta_{\Delta}(z) = \sum_{n=-\infty}^\infty \sum_{k=0}^\infty  \dd_n \l^{-z} _{n,k}
  $
  in terms of which  one finds   that  $ \zeta_{\Delta}(0)=0$  ($S^5$ is odd-dimensional)  
  and    (see, e.g., \ci{Allen:1986qi,Fursaev:2001yu,Giombi:2014yra})
  \ba \la{14}
  F=    - \tfrac{1}{2}   \zeta'_{\Delta}(0) =    F_c   + \bar F   \ , \qquad 
     F_c= \bb   E_c =   \tfrac{1}{2}  \bb \sum_{k=0}^\infty   \dd_n\,  \omega_k   \ ,\qquad 
    \bar F = \sum_{k=0}^\infty  \dd_k \log (1 - e^{- \bb \omega_k})  \ . 
  \ea
In the present case  of \rf{3.48}   we have  instead of  $ \l_{n,k}$  \rf{d0}  their "shifted"  analog 
that can be written as 
\be 
 \l_{n,k}(\nu)  = \omega^2_k  + ({2 \pi n\ov \bb}  + 2  i \nu)^2 = (\om^+_k +  i  {2 \pi n\ov \bb} )  (\om^-_k -  i  {2 \pi n\ov \bb} ) \ , 
 \qquad 
\om^\pm _k  \equiv  \om_k \pm 2 \nu = k+2 \pm 2 \nu    \ . \la{dd3} \ee 
We observe  that 
\ba
& \sum_{n=-\infty}^\infty  \log  \l_{n,k}(\nu)  = \sum_{n=-\infty}^\infty \log \big(\om^+_k +  i  {2 \pi n\ov \bb} \big)  + 
\sum_{n=-\infty}^\infty  \log \big(\om^-_k -   i  {2 \pi n\ov \bb} \big) \no \\ 
& \ \ \  = \tfrac{1}{2} \sum_{n=-\infty}^\infty \log \big[ (\om^+_k)^2  +  ({2 \pi n\ov \bb})^2 \big]
 + \tfrac{1}{2}  \sum_{n=-\infty}^\infty \log \big[ (\om^-_k)^2  +  ({2 \pi n\ov \bb})^2 \big]  \ . \la{dd4}
\ea
We can thus apply    \rf{14}   to  get  the following representation for  the free energy in \rf{3.48} 
\ba
&  F(\nu) =  F_c (\nu)  + \bar F(\nu)   =    \ha \big[  F^+(\nu)   + F^-(\nu) \big] \ , \ \ \ \ \ \ \ \qquad 
F^\pm(\nu) = F^\pm_c  + \bar F^\pm \ , \la{d6} \\
& F^\pm _c= \bb   E^\pm_c =  \ha   \bb \sum_{k=0}^\infty   \dd_k\,  \omega^\pm _k   \ ,\qquad \ \ \ \ \ \ 
    \bar F ^\pm = \sum_{k=0}^\infty  \dd_k \log \big(1 - e^{- \bb \omega^\pm_k}\big)  \ .  \la{dd6}
\ea
We thus find  that  $\bar F= \ha ( \bar F^+ + \bar F^-)$  is  given  in \rf{3.50}.  To  obtain   a  finite expression for the 
Casimir energy we  may  use the   standard "energy" zeta-function  regularization   prescription\foot{Equivalently, 
instead of $(\omega^\pm _k )^s$ one may use $\omega^\pm_k  \, e^{-\eps \omega^\pm _k}$, do the  sum and drop terms that are singular in the limit $\eps \to 0$. The expression in \rf{dd8} generalizes to $\nu\not=0$ the   value found, e.g., 
 in  \cite{Gibbons:2006ij,Beccaria:2014qea}.}
\be 
E^\pm_c (\nu) 
=   \ha   \sum_{k=0}^\infty   \dd_k\,  (\omega^\pm _k )^s\Big|_{s\to 1}  \ \to \  \ \te 
-\frac{31}{60480}+\frac{1}{18}\nu^{4}-\frac{4}{45}\nu^{6} \ . \la{dd8}
\ee
 This is even in $\nu$   so we get the expression for $E_c= \ha (E_c^++ E_c^-)$ as in \rf{3.49},\rf{351}.

\iffa 
To  compute  the  double sum  in  (\ref{3.48})
we start  with  the  corresponding  spectral zeta function  (cf.,  e.g.,   \cite{Elizalde:2002ak}) 
\be\la{dd1}
\zeta(s) = \frac{1}{24}\sum_{n\in\mathbb Z}\sum_{k=0}^{\infty}(k+1)(k+2)^{2}(k+3)\,\big[(k+2)^{2}+4(\nb+i\nu)^{2}\big]^{-s} \ ,\qquad \ \  
F= - \zeta'(0) \ . 
\ee
 Using  a  Mellin transform  $\zeta(s)$ may be written     as 
\ba
\zeta(s) &= \frac{1}{24}\sum_{n\in\mathbb Z}\sum_{k=0}^{\infty}(k+1)(k+2)^{2}(k+3)\,
\frac{1}{\Gamma(s)}\int_{0}^{\infty}dt\,t^{s-1}e^{-t\,[(k+2)^{2}+4(\nb+i\nu)^{2}]} \lp
= \frac{1}{24}\sum_{k=0}^{\infty}(k+1)(k+2)^{2}(k+3)\,\frac{1}{\Gamma(s)}\int_{0}^{\infty}dt\,t^{s-1}e^{-t\,(k+2)^{2}}
\sum_{n\in\mathbb Z}e^{-4t(\nb+i\nu)^{2}} .
\ea
Using that 
$\Theta(z,\tau) = \sum_{n\in\mathbb Z}e^{i\pi n^{2}\tau}e^{2\pi i n z} = \sqrt\frac{i}{\tau}\,\sum_{n\in\mathbb Z} e^{-\frac{i\pi}{\tau}(n+z)^{2}},
$
we may write 
\be
\sum_{n\in\mathbb Z}e^{-4t(\nb+i\nu)^{2}} = \frac{\beta}{4\sqrt{\pi t}}\sum_{\ell\in\mathbb Z}\exp\Big[-\frac{\beta^{2}}{16t}\ell^{2}+\beta \nu \ell\Big].
\ee
Separating the terms with $\ell\not=0$    and $\ell=0$ we get 
\ba
\zeta(s) =&  \zeta_{1}(s)+\zeta_{2}(s),  \qquad \qquad    \zeta(0) =0 \ ,\la{dd4}   \\  \la{dd5}
\zeta_{1}(s) =& \frac{\beta}{96\sqrt\pi} \sum_{k=0}^{\infty}(k+1)(k+2)^{2}(k+3)\,\frac{1}{\Gamma(s)}\int_{0}^{\infty}dt\,t^{s-\frac{3}{2}}e^{-t\,(k+2)^{2}}
\mathop{\sum_{\ell\in\mathbb Z}}_{\ell\neq 0}e^{-\frac{\beta^{2}}{16t}\ell^{2}+\beta\nu \ell}, \\
\zeta_{2}(s) =& \frac{\beta}{96\sqrt\pi}\frac{\Gamma(s-\frac{1}{2})}{\Gamma(s)}\sum_{k=0}^{\infty}(k+1)(k+2)^{3-2s}(k+3). \la{dd6}
\ea
To evaluate the  function $\zeta_{1}(s)$, we  need first  to  compute the following integral\footnote{This  integral converges at  $t=0$ precisely because 
we separated out the $\ell=0$ term.}
\be
h_{\ell}(s) = \int_{0}^{\infty}dt\,t^{s-\frac{3}{2}}e^{-t(k+2)^{2}-\frac{\beta^{2}}{16t}\ell^{2}} = 2^{2-2s}(k+2)^{\frac{1}{2}-s}(\beta |\ell|)^{s-\frac{1}{2}}K_{\frac{1}{2}-s}(\frac{1}{2}
|\ell|\beta(k+2)).
\ee
To find $-\zeta_{1}'(0)$  we need  only  to evaluate $h_{\ell}(0)$  since   
\be
\la{F.10}
-\partial_{s}\frac{h_{\ell}(s)}{\Gamma(s)}\Big|_{s=0} = -h_{\ell}(0) = -\frac{4\sqrt{\pi}}{|\ell|\beta}e^{-\frac{1}{2}(k+2)|\ell|\beta}.
\ee
Summing this over $\ell$ as in \rf{dd5}   gives  
\be
\la{F.11}
 -\zeta_{1}'(0) =  \frac{1}{24}\sum_{\pm} \sum_{k=0}^{\infty}(k+1)(k+2)^{2}(k+3)\,\log[1-e^{-(k+2\pm 2\nu)\frac{\beta}{2}}] \ , 
\ee
which is the expression for the "thermal"  part of free energy in \rf{3.50}.
\be
 -\zeta'_{2}(0) = \beta\frac{1}{48}\sum_{k=0}^{\infty}(k+1)(k+2)^{2}(k+3)\times (k+2) = \frac{\beta}{2} \times \frac{1}{2}\sum_{k=0}^{\infty}{\rm d}_{k} e_{k}(0).
\ee
This may be written 
\be
\la{F.13}
F_{\rm Casimir}^{S^{1}\times S^{5}} = \frac{\beta}{2} \, E_{c,\phi}(\nu), \qquad 
E_{c,\phi}(\nu) = \frac{\mc E_{c}(\nu)+\mc E_{c}(-\nu)}{2},  \qquad \mc E_{c}(\nu) = \frac{1}{2}\sum_{k=0}^{\infty}{\rm d}_{k} e_{k}(\nu).
\ee
We evaluate $\mc E_{k}(\nu)$ with a regularization factor $\mc E_{k}(\nu)^{-s}$
\ba
\mc E_{c}(\nu) = \frac{1}{24}\sum_{k=0}^{\infty}(k+1)(k+2)^{2}(k+3)(k+2+2\nu)^{s}\stackrel{s\to 1}{\to} -\frac{31}{60480}+\frac{1}{18}\nu^{4}-\frac{4}{45}\nu^{6} \ .
\ea
The value at $\nu=0$ agrees with the scalar field Casimir energy, see Eq.~(A.5) of \cite{Beccaria:2014qea} and also  \cite{Gibbons:2006ij}. 
\fi


\section{Partition functions of  fields of $(2,0)$  multiplet on  twisted $S^1 \times \wtd S^5$}
\la{app:Rtwist}

Here we shall find the  partition functions \rf{m1},\rf{4166},\rf{4177}  for   the fields  of the 6d  tensor multiplet 
defined  on  the $S^1 \times \wtd S^5$   space with the metric (cf. \rf{40},\rf{4441} and \ci{Chang:2019uag}; 
see also \ci{Bak:2016vpi})
\ba\la{e1} 
&ds^{2}_{S^{1}\times \wtd S^{5}} =\tfrac{1}{4}  dy^{2}+dn_{1}^{2}+dn_{2}^{2}+dn_{3}^{2}+n_{1}^{2}(d\varphi_{1}+i\alpha dy)^{2}
+n_{2}^{2}(d\varphi_{2}-i\alpha dy)
+n_{3}^{2}d\varphi_{3}^{2} \ ,\\
& \qquad \ \ \  y \in (0, \b), \ \ \ \  \qquad \qquad 
    n_1^2 + n_2^2 + n_3^2 =1  \ . \la{e2}
\ea
We shall start with  computing the  conformally coupled 
scalar  free energy directly from the determinant of  its kinetic operator and  also 
 show   how to obtain the same result by the operator counting method  in "twisted" $\mathbb R^{6}$.
 We shall then apply the latter approach to the fermion and the antisymmetric tensor cases.




\def \half {{\frac{1}{2}}}

\subsection{Scalar free energy on $S^1 \times \wtd S^5$}

Let us  first  consider  the  simplest  case of a scalar  on  $S^{1}_{\b}\times \wtd S^{1}$
with the metric ${1\ov 4} d y^{2}+(d\varphi+i\alpha dy)^{2}$, where $\varphi \in (0, 2\pi)$. 
Introducing $\vp'= \varphi+i\alpha y$ one finds that  
$-\nabla^{2} = -4\partial_{y}^{2}-\partial_{\varphi'}^{2}$  has  the eigenfunctions
$f_k(p)\, \exp( i p  y+ik\varphi') = \exp[i(p  +i\alpha k)\,y+ik\varphi]$  which are  periodic under 
$y\to y + \b$ and $\varphi \to \varphi + 2 \pi$   if $p  +i\alpha k = \nb = {2 \pi n\ov \b}$ 
 where $n$ and $k $ are integers.  The corresponding eigenvalue is $4p^2 + k^2 = 4 (\nb -i \a k)^2 + k^2$, i.e. 
\be\la{e4}
F = \tfrac{1}{2}\log\det(-\nabla^{2}) = \tfrac{1}{2}\sum_{n,k\in \mathbb Z}\log \big[4 (\nb-i\alpha k)^{2}+k^{2}\big] \ . 
\ee
The sum over $n$   here can be done as in  section  \ref{sec:free-sc} or Appendix \ref{app:free-energy}
with the shift $\nu = \alpha k$. We find 
for the thermal part of the free energy
\be\la{e5}
\bar F = \tfrac{1}{4}\sum_{\pm}\sum_{k=1}^{\infty}\log(1-e^{- {1\ov 2}  \beta k(1\pm2 \alpha)}).
\ee
The corresponding 
 single-particle partition function is then (cf. \rf{444})
\be
\la{5.8}
 \Zsp_{\phi}(q,u) = 
 \sum_{\pm}\sum_{k=1}^{\infty}q^{k(1\pm 2\alpha)} = \sum_{\pm}\frac{q^{1\pm 2\alpha}}{1-q^{1\pm2\alpha}} = \sum_{\pm}\frac{u^{\pm 1}q}{1-u^{\pm 1}q} \ ,  \ \ \ \ \ \qquad
 q= e^{-\half \beta}, \ \ \ \    u= q^{2 \a} \ . 
\ee
Next, let us consider a  more complicated case of $S^{1}_{\b}\times \wtd S^{3}$  subset of \rf{e1} 
with the metric   ${1\ov 4}  dy^{2}+d\chi^2 +\cos^2 \chi\, (d\varphi_{1}+i\alpha dy)^{2}
+\sin^2 \chi\, (d\varphi_{2}-i\alpha dy)^2$.
A convenient basis for the eigenfunctions  of the Laplacian on the standard  $S^3$ is  (see, e.g., \cite{lehoucq2003eigenmodes,lachieze2004laplacian})
\be
\la{5.19}
\Phi_{k, r_{1},r_{2}} = F_{k, r_{1}, r_{2}}(\chi)\ e^{i(r_{1}+r_{2})\vp_{1}}\ e^{i(r_{2}-r_{1})\vp_{2}},\qquad \qquad 
r_{1},r_{2}=-\ha {k}, \dots, \ha{k} \ , \ \ \   k=0,1, ...  \ . 
\ee
Including  the  effect of the two  $\alpha$-shifts  is then  achieved  as in the above example  
and we find  the following analog of  \rf{e4}  for a  conformally coupled massless scalar\foot{For the standard Laplacian on $S^3$ 
 we have  $\lambda_k = k(k+2)$    and $\dd_k= (k + 1)^2 $ 
 and for a conformally coupled scalar  we need to add 1 to $\lambda_k $
 (see  footnote \ref{foot:lapl}).}
 \ba
F = \tfrac{1}{2}\log\det(-\nabla^{2}  +1)& = \tfrac{1}{2}\sum_{k=0}^{\infty}\, \sum_{r_{1},r_{2}=-k/2}^{k/2}
\log \Big[4 \big(\nb+i\alpha (r_{1}+r_{2})-i\alpha(r_{1}-r_{2})\big)^{2}+k(k+2)+1\Big]\lp
=  \tfrac{1}{2}\sum_{k=0}^{\infty}\sum_{m=-k/2}^{k/2}(k+1)
\log \big[4(\nb+2i\alpha m)^{2}+(k+1)^{2}\big]\ .  \la{e7}
\ea
The  corresponding  thermal partition function   and the single-particle partition function  are then 
\ba
&\qquad \qquad  \bar F =  \tfrac{1}{4}\sum_{\pm}\sum_{k=0}^{\infty}\sum_{m=-k/2}^{k/2}(k+1)
\log[1-e^{-\ha \beta(k+1\pm 4\alpha m)}]  \ , \\
\la{5.22}
&S^{1}\times \wtd S^{3}: \qquad \Zsp_{\phi}(q,u) =  \tfrac{1}{2}\sum_{\pm}\sum_{k=0}^{\infty}\sum_{m=-k/2}^{k/2}(k+1)q^{k+1\pm 4\alpha m} = \frac{q-q^{3}}{(1-q^{1+2\alpha})^{2}(1-q^{1-2\alpha})^{2}} \notag \\
&\qquad\quad\ \ \qquad\quad \qquad \quad =  \frac{q(1-q^{2})}{(1-u\,q)^{2}(1-u^{-1}\,q)^{2}} \ , \ \ \ \ \ \ \ \     q= e^{-\half \beta}, \ \ \ \    u= q^{2 \a} \ . 
\ea
The computation in  the case of $S^{1}_{\b}\times \wtd S^{5}$  we are interested in   is similar. 
%
One  is first  to  find  a convenient labelling of the eigenfunctions of the Laplacian  on $S^5$ 
represented by  
 $k$-symmetric traceless tensors built out of Cartesian coordinates of $\mathbb R^6$. More precisely, we need
the weights of the $k$-symmetric traceless representation ${\rm S}_{k}$ of $SO(6)$.

The solution to this problem is based on the  embedding 
$
SU(2) \times SU(2) \times U(1)\subset SO(6).
$
The  15 generators of $SO(6)$ are $J_{mn}$ with $m,n=1, \dots, 6$  with 
the Cartan subalgebra generators $J_{12}, J_{34}, J_{56}$.
 The two $SU(2)$'s   are the factors in $SO(4)=SU(2)\times SU(2) $ where the generators
of $SO(4)$ are $J_{ab}$ with $a,b=1, \dots, 4$. The generator of $U(1)$  is  $J_{56}$. Thus the  states in a generic  
representation of $SO(6)$ are 
$
|j, j'; m, m'; s\rangle,\  m = -j, \dots, j, \ m' = -j', \dots, j',
$
where $j, j'$ are integer or half-integer, $m,m'$ have unit spacing, and the  eigenvalue $s$ of $J_{56}$  is integer. 
For the ${\rm S}_{k}$ representation of $SO(6)$ one has 
$j'=j$.

 All states are degenerate with respect to the  ladder operators of the two  $SU(2)$'s, \ie $J_{\pm}$ and $J'_{\pm}$. Thus we can restrict our attention to the states 
$
|j, j; j, j; s\rangle
$
that contribute to the spectrum with multiplicity $(2j+1)^{2}$. As shown in \cite{piepenbring19872},  such  states appearing in ${\rm S}_{k}$ have
quantum numbers 
\be\te 
j = 0, \frac{1}{2}, 1, \frac{3}{2}, \dots, \frac{k}{2}, \qquad -(k-2j)\le s \le k-2j, \ \ \ \ s\  \text{in steps of 2}.
\ee
The number of $s$ values  for a given $j$ is thus $k-2j+1$,  reproducing the multiplicity ${\rm d}_{k}$
of the  eigenvalues of the Laplacian on $S^5$ 
\be
\sum_{j=0, 1/2, \dots, k/2}(k-2j+1)(2j+1)^{2} = \sum_{n=0}^{k}(k-n+1)(n+1)^{2} = \tfrac{1}{12}(k+1)(k+2)^{2}(k+3)\ .
\ee
The free energy for a  conformally coupled massless scalar  on $S^1 \times  \wtd S^{5}$ 
 is then found to be\footnote{We use the explicit relation between the two $SU(2)$ generators
and the $SO(4)$ generators $J_{3} =\frac{1}{2}( J_{12}+J_{34})$,
$J_{3}' =\frac{1}{2}( J_{12}-J_{34})$.} 
\ba
\la{5.39}
F &= \tfrac{1}{2}\log\det(-\nabla^{2} + 4)  \lp =
\tfrac{1}{2}\sum_{k=0}^{\infty}\, \sum_{j=0,1/2,\dots,k/2}(k-2j+1)\sum_{r_{1},r_{2}=-j}^{j}
\log \Big[4\big(\nb+i\alpha (r_{1}+r_{2})-i\alpha(r_{1}-r_{2})\big)^{2}+(k+2)^{2}\Big]\lp
=  \tfrac{1}{2}\sum_{k=0}^{\infty}\, \sum_{j=0,1/2,\dots,k/2}(k-2j+1)(2j+1)\sum_{r_2=-j}^{j}
\log \big[2(\nb+2i\alpha r_2)^{2}+(k+2)^{2}\big].
\ea
The corresponding single particle partition function reads
\ba
\la{5.400}
S^{1}\times \wtd S^{5}: \qquad \Zsp_{\phi}(q,u) & =\tfrac{1}{2}\sum_{\pm}\sum_{k=0}^{\infty}\, \sum_{j=0,1/2,\dots,k/2}(k-2j+1)(2j+1)\sum_{m=-j}^{j}q^{k+2\pm 4\alpha m} \lp
= \frac{q^{2}(1-q^{2})}{(1-q)^{2}(1-q^{1+2\alpha})^{2}(1-q^{1-2\alpha})^{2}} = 
\frac{q^{2}(1-q^{2})}{(1-q)^{2}(1-u\,q)^{2}(1-u^{-1}q)^{2}}\ .
\ea
As in (\ref{5.22}),  here the only dependence on $u$ is in the factors in  the denominator.
Eq.\rf{5.400}  thus  reproduces the expression in \rf{m1}.

\subsection{Single-particle partition functions from operator counting  on  twisted $\mathbb R^{6}$}

Let us   note that  the same   results can be found  by first mapping the conformal   scalar on   $S^{1}\times \wtd S^{d}$
to flat $\mathbb R^{d+1}$  space and then applying the operator counting method 
(see, e.g., \cite{Cardy:1991kr,Kutasov:2000td,Beccaria:2014jxa}).\foot{To recall,  one has 
$ (d\log r)^{2}+dS_{5} = \frac{1}{r^{2}}ds^{2}_{\mathbb R^{6}}$
so that  the  energies in $S^{1}\times S^{5}$ may be identified with the scaling dimensions in $\mathbb R^{6}$.} 
In the untwisted case $\alpha=0$   one starts with a free  massless  scalar 
in  $\mathbb R^{6}$ that has  dimension 2. This   corresponds to the   factor $q^2$ in \rf{5.400}. The
 factor $1/(1-q)^{6}$ counts  the number  of 
 descendant  fields obtained  by applying, in all possible ways,  the    six  $\partial_{m}$  
 derivatives to the scalar $\phi$.
The subtraction  of $q^{4}$ term  in \rf{5.400} accounts for the fact that 
 the  trivial operators proportional to the  equations of motion $\partial^{2}\phi$  should not be included. 
 Similar    interpretation is true for  \rf{5.22}. 
 
 To include the twists  one  may  separate   coordinates  into 1+2  rotation  planes   and multiply the relevant derivatives 
  by the corresponding  rotation phases.
    The numerator in \rf{5.400} is then unchanged as the equation of motion is rotationally invariant. 
    The same   applies to \rf{5.22}.\foot{Note that 
    in the  case of $S^{1}\times \wtd S^{1}$ with only one twist the expression in \rf{5.8} 
    may  be also  interpreted  in terms of the  operator counting  on $\mathbb R^2$ 
     as the contribution of  one tower of fields  starting with dimension 1  primary 
     $\del_z \phi$ and  another corresponding to its conjugate $ \del_{\bar z} \phi$, 
          with the twist introducing factors of $u$ and $u^{-1}$   respectively. }


This  operator counting  method  in twisted $\mathbb R^{6}$ can
 be applied also  to  find the single-particle partition functions the  fermion \rf{4166}
 and 
 the antisymmetric tensor  \rf{4177}. 
  For the latter  with the  field  strength $H_{mnk}$ of dimension 3 
  in the  untwisted  case  one has \cite{Kutasov:2000td,Beccaria:2014jxa}
 \be
\la{5.43}
\Zsp_{A}(q) =  \frac{10q^3 -15q^4+6q^{5}-q^{6}}{(1-q)^{6}}=\frac{q^{3}(10-5q+q^{2})}{(1-q)^{5}} \ . 
\ee
Here the coefficient 10 of the  $q^{3}$  term   in the numerator 
  corresponds to the  number  $\ha \times {6\cdot 5\cdot 4\ov 3!} = 10$ of  the 
$H_{mnk}$ components subject to  the self-duality condition. 
The  two $\alpha$-twists   are  taken into account by assigning to the six derivatives $\del_m$   of $\mathbb R^{6}$
    the $R$-charge weights
\be
\la{5.44}
r_{k} = (u,u,u^{-1},u^{-1},1,1) \ , 
\ee
corresponding to the opposite  rotations  in  the 12 and 34  planes (cf. \rf{e1}). 
Then, before imposing the self-duality,  the 20 components of $H_{mnk}$ have  the following   rotation weights  
\be
\def\arraystretch{1.3}
\begin{array}{cc}
\toprule
\text{3-form components} & \text{weight} \\
\midrule
H_{125}, H_{126} & 2\times u^{2} \\
H_{123}, H_{124}, H_{156}, H_{256} & 4\times u \\
H_{135}, H_{136}, H_{145}, H_{146}, H_{235}, H_{236}, H_{245}, H_{246} & 8\times u^{0}\\
H_{134}, H_{234}, H_{356}, H_{456} & 4\times u^{-1}\\
H_{345}, H_{346}& 2\times u^{-2}\\
\bottomrule
\end{array}\notag
\ee
As a result, the $10 q^3$ term  in \rf{5.43}  gets    generalized   to 
\be
10q^{3}\ \to\  c_1(u)\, q^{3}, \qquad \qquad  c_1(u) = u^{2}+2u+4+2u^{-1}+u^{-2}.
\ee
The term  $-15q^{4}$  in \rf{5.43}  corresponds to the subtraction of 
the  contribution of the equation of motion
  operator\foot{The self-duality constraint implies that $\partial^{m}H_{mnk}^{\star}$  should not be considered.} 
   $\partial^m H_{mnk} \equiv  \tilde H_{nk}$  
  for which 
similar  
 assignments of  the  twist  factors read
 \be
\def\arraystretch{1.3}
\begin{array}{cc}
\toprule
\text{$\tilde H_{mn}$ components} & \text{weight} \\
\midrule
\tilde H_{12} & u^{2} \\
\tilde H_{15}, \tilde H_{16}, \tilde H_{25}, \tilde  H_{26} & 4\times u \\
\tilde H_{13}, \tilde H_{14}, \tilde H_{23}, \tilde H_{24}, \tilde H_{56} & 5\times u^{0}\\
\tilde H_{35}, \tilde H_{36}, \tilde H_{45}, \tilde H_{46} & 4\times u^{-1}\\
\tilde H_{34}& u^{-2}\\
\bottomrule
\end{array}\notag
\ee
As a result, we get 
\be
-15q^{4} \ \to\   -(u^{2}+4u+5+4u^{-1}+u^{-2})\,q^{4} = -[c_1(u)+c_2(u)]\,q^{4}, \qquad c_2(u) = 2u+1+2u^{-1}.
\ee
One concludes  that (\ref{5.43}) is to be  replaced by 
\be
\la{5.47}
\Zsp_{A}(q,u)=\frac{q^{3}[c_1(u)-c_2(u)\,q+q^{2}]}{(1-q)(1-u\,q)^{2}(1-u^{-1}q)^{2}} \ , 
\ee
which reproduces \rf{4177}.

Similarly, in the  case of the MW  6d  fermion  of dimension  ${5\ov 2}$  one finds  the expression in \rf{4166}, i.e. 
\ba
\la{5.48}
\Zsp_{\psi}(q) = \frac{2q^{5/2}(1-q) }{(1-q)^{6}}  
\to  \Zsp_{\psi}(q,u) = \frac{2c_{\psi}(u)\,q^{5/2}}{(1-q)(1-u\,q)^{2}(1-u^{-1}q)^{2}},
 \qquad c_{\psi}(u) = \tfrac{1}{4}(u+2+u^{-1}). 
\ea
Here in addition  to the  twist  factors for the derivatives  leading to the $u$ dependence in the denominator  we get  also 
the factor $c_{\psi}(u) = \frac{1}{4}(u+1+1+u^{-1}) =\frac{1}{4} (u^{1/2}+u^{-1/2})^{2}$  
that accounts for the rotation weights  of  different components of   $\psi$. 
The effect of rotation in the two planes  is represented by 
 the matrix $\exp(\frac{i}{2} \delta \varphi_1\Gamma_{12}+\frac{i}{2}\delta \varphi_2\Gamma_{34})$.
 It  can be diagonalized to give the factor $\exp\big[\frac{1}{2}(\pm \varphi_1\pm \varphi_2)\big]$ with four independent
  sign combinations.
For two opposite twists one has  $\delta \varphi_1 = -\delta \varphi_2
= \beta \alpha$ and this  leads to  the sum of the terms  $u + 2+  u^{-1}$  in $c_{\psi}(u)$.


Following the same logic, it is straightforward 
also 
to write down  the partition functions 
 for the case of the  two independent twists $\alpha$ and $ \alpha'$  in the two  isometric   angles  in \rf{e1}.  
Generalizing  (\ref{5.39}) and (\ref{5.400}), for the scalar field 
 we get (here $u=q^{2\alpha}$ and $u'=q^{2\alpha'}$)
\ba
\Zsp_{\phi}(q, u, u')  =& \tfrac{1}{2}\sum_{\pm}\sum_{k=0}^{\infty}\sum_{j=0,1/2,\dots,k/2}(k-2j+1)\sum_{r_{1},r_{2}=-j}^{j}q^{k+2 \pm (2\alpha(r_{1}+r_{2})+2\alpha'(r_{1}-r_{2}))} \lp
= \frac{q^{2}(1-q^{2})}{(1-q)^{2}(1-q^{1+2\alpha})(1-q^{1-2\alpha})(1-q^{1+2\alpha'})(1-q^{1-2\alpha'})}.
\ea
For the tensor field  one has to generalize (\ref{5.44}) to $r_{k} = (u,u',u^{-1},u'^{-1},1,1)$ .
For the fermion, considering two independent twists gives \rf{5.48}  with 
$c_{\psi}(u,u') = \frac{1}{4}(u^{1/2}+u^{-1/2})(u'^{1/2}+u'^{-1/2})$.

\


\small 

\bibliography{BT-Biblio}
\bibliographystyle{JHEP-v2.9}
\end{document}